\documentclass[aps,preprint,showpacs]{revtex4}
 \usepackage{epsfig}
 \usepackage{isolatin1}

 \usepackage{makeidx} \makeindex

 \usepackage{amsmath} \usepackage{amssymb} \usepackage{amsthm}

 \usepackage{mathrsfs} 

 \def\id{1\hskip-2.5pt{\rm l}}

\begin{document}

 \title {Bang-bang refocusing of a qubit exposed to telegraph noise}


 \begin{abstract}
 Solid state qubits promise the great advantage of being naturally
 scalable to large quantum computer architectures, but they also
 possess the significant disadvantage of being intrinsically
 exposed to many sources of noise in the macroscopic solid-state
 environment. With suitably chosen systems such as superconductors,
 many of sources of noise can be suppressed. However, imprecision
 in nanofabrication will inevitably induce defects and disorder,
 such as charged impurities in the device material or substrate.
 Such defects generically produce telegraph noise and can hence be
 modelled as bistable fluctuators. We demonstrate the possibility
 of the active suppression of such telegraph noise by bang-bang
 control through an exhaustive study of a qubit coupled to a single
 bistable fluctuator.  We use a stochastic Schr\"odinger equation,
 which is solved both numerically and analytically. The resulting
 dynamics can be visualized as diffusion of a spin vector on the
 Bloch sphere.

 We find that bang-bang control suppresses the effect of a bistable
 fluctuator by a factor roughly equalling the ratio of the
 bang-bang period and the typical fluctuator period. Therefore, we
 show the bang-bang protocol works essentially as a high pass
 filter on the spectrum of such telegraph noise sources. This
 suggests how the influence of $1/f$-noise ubiquitous to the solid
 state world could be reduced, as it is typically
 generated by an ensemble of bistable fluctuators.  Finally, we
 develop random walk models that estimate the level of noise
 suppression resulting from imperfect bang-bang operations, such as
 those that cannot be treated as $\delta$-function impulses and
 those that have phase and axis errors.

 \end{abstract}

 \author{Henryk Gutmann} \affiliation{Sektion Physik and CeNS,
 Ludwig-Maximilians-Universit\"at, 80333 M\"unchen, Germany}
 \author{William M. Kaminsky} \affiliation{Department of Physics,
 Massachusetts Institute of Technology, Cambridge, MA 02139}
 \author{Seth Lloyd} \affiliation{Department of Mechanical Engineering,
 Massachusetts Institute of Technology, Cambridge, MA 02139}
 \author{Frank K. Wilhelm}  \affiliation{Sektion Physik and CeNS,
 Ludwig-Maximilians-Universit\"at, 80333 M\"unchen, Germany}

 \email[email: ]{gutmann@theorie.physik.uni-muenchen.de}


 \maketitle

 \section{Introduction}

 In order to implement solid-state quantum information processing
 devices, the noise sources causing decoherence of their quantum
 states have to be carefully understood, controlled, and
 eliminated. This is a formidable task, as a solid-state
 environment generically couples a macroscopic number of degrees of
 freedom to any such device. Thus, a fundamental prerequisite for
 any design is that it must significantly decouple the quantum
 states used for computation from phonons and other quasiparticles
 in the underlying solid crystal.  Examples of such designs are
 those employing discrete states in quantum dots \cite{LDV98} or
 superconductors with a gapped density of states \cite{Vion,Chiorescu03,Martinis02,Nakamura99,Makhlin01}.

 Most research going beyond this fundamental prerequisite has
 concentrated on decoupling devices from external noise sources
 such as electromagnetic noise generated by control and measurement
 apparatus \cite{EPJB}. On the other hand, there inevitably are
 internal noise sources because the fabrication of gates, tunnel
 junctions, and other functional components creates defects in the
 underlying crystal. Prominent examples of such defects 
are background charges in charge-based devices or cricital current 
fluctuations in flux-based devices \cite{Harlingen,ZCC97}. A clear signature of such defects is telegraph
 noise in the case of a few defects or $1/f$-noise in the case of a
 larger ensemble \cite{Weissmann}. With the growing success
 in engineering the electromagnetic environment, these defects are
 becoming more and more the key limiting sources of decoherence.

 Such defects do not fall in the large class of noise sources that
 can be approximated well as a bosonic bath, and this fact
 complicates analysis. Whereas it is realistic to treat a bosonic
 bath in the tractable near-equilibrium thermodynamic limit where
 fluctuations are purely Gaussian \cite{WeissQDS,Leggett87,SMS02},
 localized noise sources with bounded spectra like the defects in
 which we are interested produce noise that is significantly
 non-Gaussian.  Theories treating large ensembles of non-Gaussian
 noise sources have been presented \cite{GMB02,Stamp}.  However,
 with the ongoing improvement in nanofabrication technology, it is
 realistic to consider the case where non-Gaussian noise sources
 are reduced down to only a single one or a few per device.  This
 is the case we treat here, and thus the defects find a more
 realistic representation as a small set of bistable fluctuators
 \cite{TIYT03} (henceforth abbreviated bfls).  In principle, this
 approach can be extended to larger sets of bfls with a range of
 different mean switching times (\textit{e.g.}, an ensemble with an
 exponential distribution of switching times that produces
 $1/f$-noise \cite{D&H,Martinis,Altshuler}).

 This report is organized as follows. Section II presents the model
 of a single bfl in the semiclassical limit, where it acts as a
 source of telegraph noise.  Section III introduces an idealized
 open loop quantum control technique, quantum bang-bang control
 \cite{L&V98,LVK99a,LVK99b}, which is suitable for slowly
 fluctuating noise sources.  Section IV explains how we simulated
 the qubit dynamics under the influence of noise with and without
 bang-bang control by integrating of the corresponding
 time-dependent Schr\"odinger equation. As a measure of the
 decoherence, we analyze the deviations of the qubit's trajectory
 on the Bloch sphere from that of the noiseless case.  These
 deviations take the form of a random walk around the
 noiseless-case trajectory.  We therefore analyze the suppression
 of these deviations by comparing the variances of these random
 walks with and without bang-bang control. Both numerical and
 analytical solutions (the latter in the long-time or ``diffusion''
 limit) are presented. Comparison of the numerical simulations to
 the analytical solutions shows excellent agreement. We then
 analyze how these results change when practical limitations are
 considered such as the fact that a bang-bang pulse cannot be an ideal
 $\delta$-function impulse and the fact that the duration or
 polarization axis of the pulse may suffer from random fluctuations. 
 We show at the end of Section IV.B that within large margins
 bang-bang suppression of the bfl noise is not inhibited by having
 a finite, rather than infinitesimal, pulse length. However, in
 Section V, we do find that duration and polarization axis errors
 in the bang-bang pulses can significantly affect the suppression
 of bfl noise. We present a point of optimum performance. Section
 VI concludes with remarks on several recent publications
 concerning the suppression of telegraph or $1/f$-noise.

 \section{Model of the bistable fluctuator in its semiclassical limit}

\begin{figure}[h]
 \includegraphics[width=0.99\columnwidth]{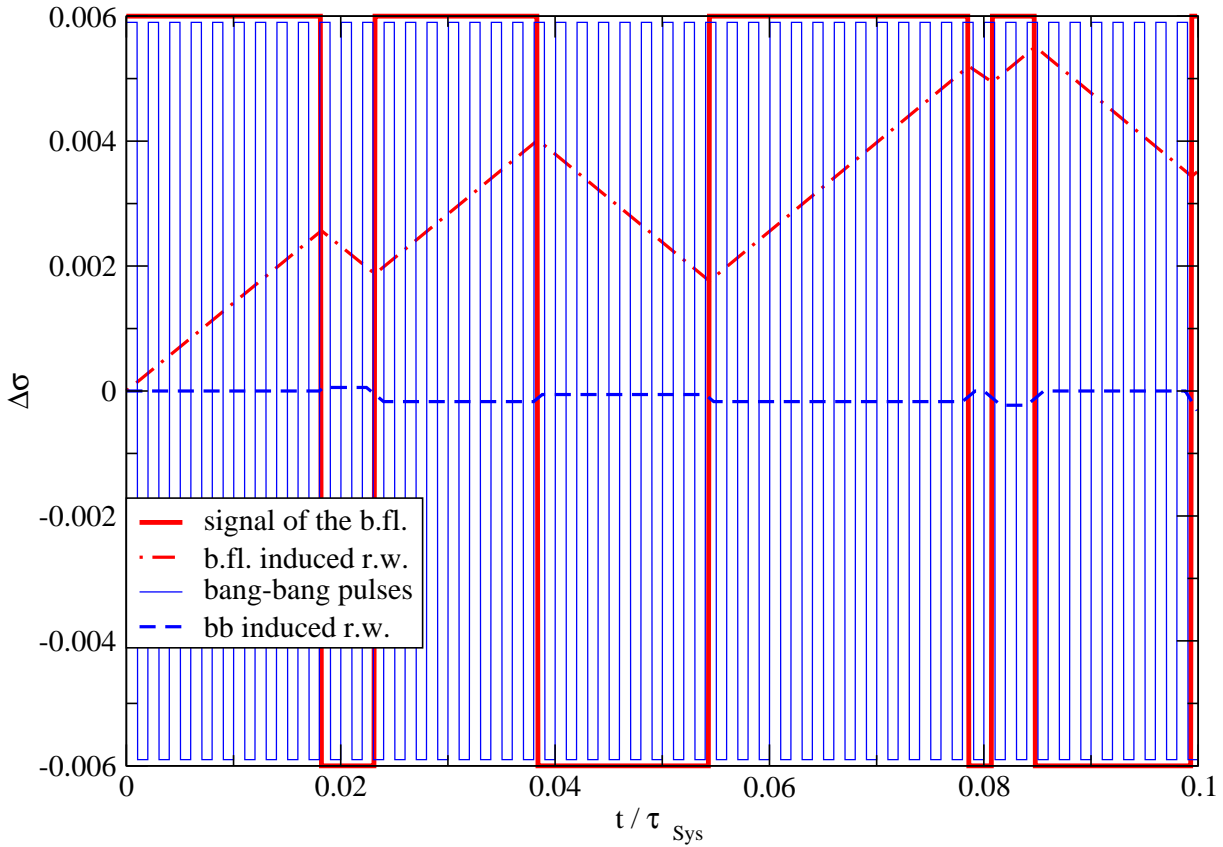}
 \caption{\label{bflnoisesignal}Schematic plot of a typical
 Poisonian bfl noise signal and its resulting random walk behavior
 (in the limit of small deviations). The periodic fast switching
 step function represents a bang-bang pulse with a time scale
 ratio: $\tau_{\rm bfl}/\tau_{\rm bb} = 10$ and yields a quite
 smaller random walk step-length. $\tau_{\rm Sys}={\pi \over
 \sqrt{\epsilon_{\rm q}^2+\Delta_{\rm q}^2}}$ denotes the evolution
 period of the qubit in the noiseless case.}
 \end{figure}

 We describe the bfl-noise influenced evolution of the qubit in its
 semiclassical limit by using a stochastic Schr\"odinger equation
 \cite{vKp,Arn73} with the time-dependent effective Hamiltonian
 \begin{eqnarray}\label{Hstocheq}
 H_{q}^{\rm eff}(t)&=&H_{\rm q}+H_{\rm noise}(t)\\
 \label{statham}H_{\rm q}&=&\hbar \epsilon_{\rm q} \hat{\sigma}_{\rm z}^{\rm q} +
 \hbar \Delta_{\rm q} \hat{\sigma}_{\rm x}^{\rm q}\\
 H_{\rm noise}(t)&=&\hbar
 \alpha~\hat{\sigma}_{\rm z}^{\rm q}~\xi_{\rm bfl}(t)
 \end{eqnarray}
 where $\epsilon_{\rm q}$ and $\Delta_{\rm q}$ define the free
 (noiseless) qubit dynamics. $\xi_{\rm bfl}(t)$ denotes a function
 randomly switching between $\pm1$ (see Fig.~\ref{bflnoisesignal}),
 which represents a telegraph noise signal. The switching events
 follow a symmetrical Poisson process, \textit{i.e.}, the
 probabilities of the bfl switching from $+1$ to $-1$ or $-1$ to
 $+1$ are the same and equal in time. The Poisson process is
 characterized by the mean time separation $\tau_{\rm bfl}$ between
 two bfl flips. The coupling amplitude to the qubit in frequency
 units is $\alpha$. The relation of this Hamiltonian to a
 microscopic model is explained in the Appendix.\\
~~\\

 Starting with an arbitrary initial state of the qubit, represented
 by some given point on the Bloch sphere, we can numerically
 integrate the corresponding stochastic differential equation and
 obtain the corresponding random walk on the Bloch sphere
 \begin{eqnarray}
 \label{ssesol}
 \vec{\sigma}(t) & = & {\rm T} \exp \left({-i}/{\hbar} \int_0^t
 H_{\rm q}^{\rm eff} (s)\,ds  \right) \vec{\sigma}(0)
 \end{eqnarray}
 with ${\rm T}$ denoting the usual time-ordering operator.

 \begin{figure}[h]
 \includegraphics[width=0.99\columnwidth]{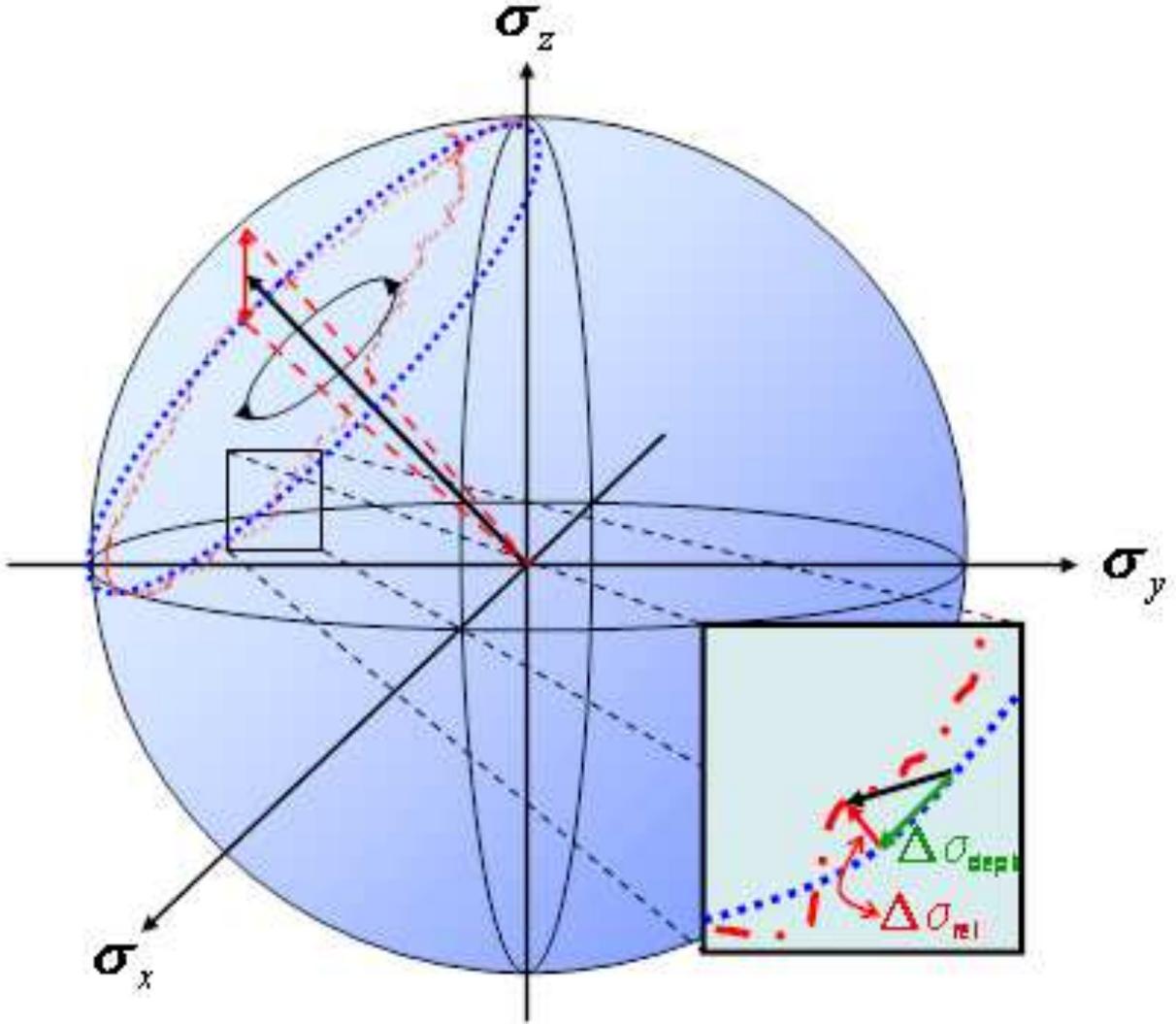}
 \caption{\label{blochsphere}Schematic plot of noisy qubit evolution
 generated by Poissonian telegraph noise.  The resulting random
 walk (dot-dashed line) on the Bloch sphere is comprised both of deviations 
 $\Delta \sigma_{\rm deph}$ in parallel to the free precession trajectory 
 (dotted line), which correspond to dephasing, and deviations 
 $\Delta \sigma_{\rm rel}$ perpendicular to it, which correspond to
 relaxation/excitation.}
 \end{figure}

 \section{Bang-bang control protocol}

 We propose to reduce the influence of the bfl-noise by applying to
 the qubit a continuous train of $\pi$-pulses along the
 $\sigma_{\rm x}$-axis.  This refocusing pulse scheme essentially
 corresponds to the standard quantum bang-bang procedure
 \cite{L&V98,LVK99a,LVK99b} or the Carr-Purcell-Gill-Meiboom echo
 technique from NMR \cite{CarrPurcell}.  For technical
 convenience, we consider the $\pi$-pulses to be of infinitesimal
 duration.  This simplification is not crucial as will be detailed
 later in Section IV.B. The pulses are assumed to be separated by a
 constant time interval $\tau_{\rm bb}$.  The mean separation
 $\tau_{\rm bfl}$ between two bfl flips is assumed to be much longer
 than $\tau_{\rm bb}$.  For theoretical convenience, we also assume
 that $\tau_{\rm bfl}$ is shorter than the free precession period
 of the qubit.  This too is not a crucial restriction. (It can
 always be overcome by changing to a co-precessing frame.)

 Qualitatively, bang-bang control works as follows. Since
 $\tau_{\rm bb} \ll \tau_{\rm bfl}$, it is usually the case that
 the bfl does not flip during the time between two bang-bang pulses
 that flip the qubit.  In this way, the bang-bang pulses average
 out the influence of $H_{\rm noise}(t)$. In fact, the refocusing 
 scheme fully suppresses the $\sigma_z$-term of the static Hamiltonian 
 (\ref{statham} (compare Fig.~\ref{bb1rwstep}); but this turns out to be 
 no crucial obstacle to universal quantum computation as will be
outlined later on.
 As one can visualize in
 Fig.~\ref{bflnoisesignal}, it is only when a bfl flip occurs
 during a bang-bang period that the net influence of the bfl felt
 by the qubit is nonzero, and the qubit thus suffers some random
 deviation from its trajectory in the noiseless case. Taken
 together, these random deviations constitute a random walk around
 the noiseless case trajectory.  While this walk is actually
 continuous, it can be modelled as a discrete walk with steps that
 are randomly distributed in time, one step for each bfl flip (see
 \textit{e.g.}~\cite{WeissRW}). The average step length is
 essentially the product of the noise coupling strength $\alpha$
 and the mean time the bfl in its present state can influence the
 qubit. Without bang-bang control, this mean influence time is
 $\tau_{\rm bfl}$, whereas with bang-bang control, it is reduced to
 $\tau_{\rm bb}$. Therefore, both with and without bang-bang
 control, the random walk has the same time distribution of steps,
 but with bang-bang control the step size can be significantly
 reduced roughly by a factor of the ratio of time scales $\tau_{\rm
 bb} / \tau_{\rm bfl}$.

 \section{Random Walk on the Bloch sphere}

 Now we study this proposal quantitatively.  We simulate these
 random walks both with and without bang-bang control by
 integrating both numerically and analytically the Schr\"odinger
 equation, Equ.~(4), with the stochastic Hamiltonian of
 Equs.~(1-3). As generic conditions for the qubit dynamics, we
 choose $\epsilon_{\rm q} = \Delta_{\rm q} \equiv \Omega_0$.
 Without loss of generality, we set the qubit's initial state to be
 spin-up along the $z$-axis. If the qubit-bfl coupling $\alpha$
 were zero, then the qubit would simply precess freely on the Bloch
 sphere around the rotation axis $\hat{\sigma}_{\rm x}^{\rm
 q}+\hat{\sigma}_{\rm z}^{\rm q}$ (the dotted line in
 Fig.~\ref{blochsphere}).  Hence, we expect for a sufficiently
 small coupling ($\alpha \ll \Omega_0$) only a slight deviation of
 the individual time evolution compared to the free evolution case
 (the dashed line in Fig.~\ref{blochsphere}).  For the coupling
 strength, we take $\alpha=0.1 \Omega_0$. All the following times
 and energies are given in units of the unperturbed system
 Hamiltonian, \textit{i.e.}, our time unit $\tau_{\rm Sys}$ is
 given according the free precession time $\pi \tau_{\rm
 Sys}/\sqrt{2}$, and our energy unit is given by $\Delta E =
 \sqrt{\epsilon_{\rm q}^2 + \Delta_{\rm q}^2} = \sqrt{2} \Omega_0$. The time
 scale ratio is taken to be $\tau_{\rm bfl}/\tau_{\rm bb}=10$ if
 not denoted otherwise.

 This approach accounts for the essential features of our specific
 situation: the long correlation time of the external noise, essentially
$\tau_{\rm bfl}$, its
 non-Gaussian statistics and its potentially large amplitude at low
 frequencies. These properties are crucial and are difficult,
 although not impossible, to take into account in standard master
 equation methods.

 \subsection{Numerical simulations}

\begin{figure}[h]
 \includegraphics[width=0.99\columnwidth]{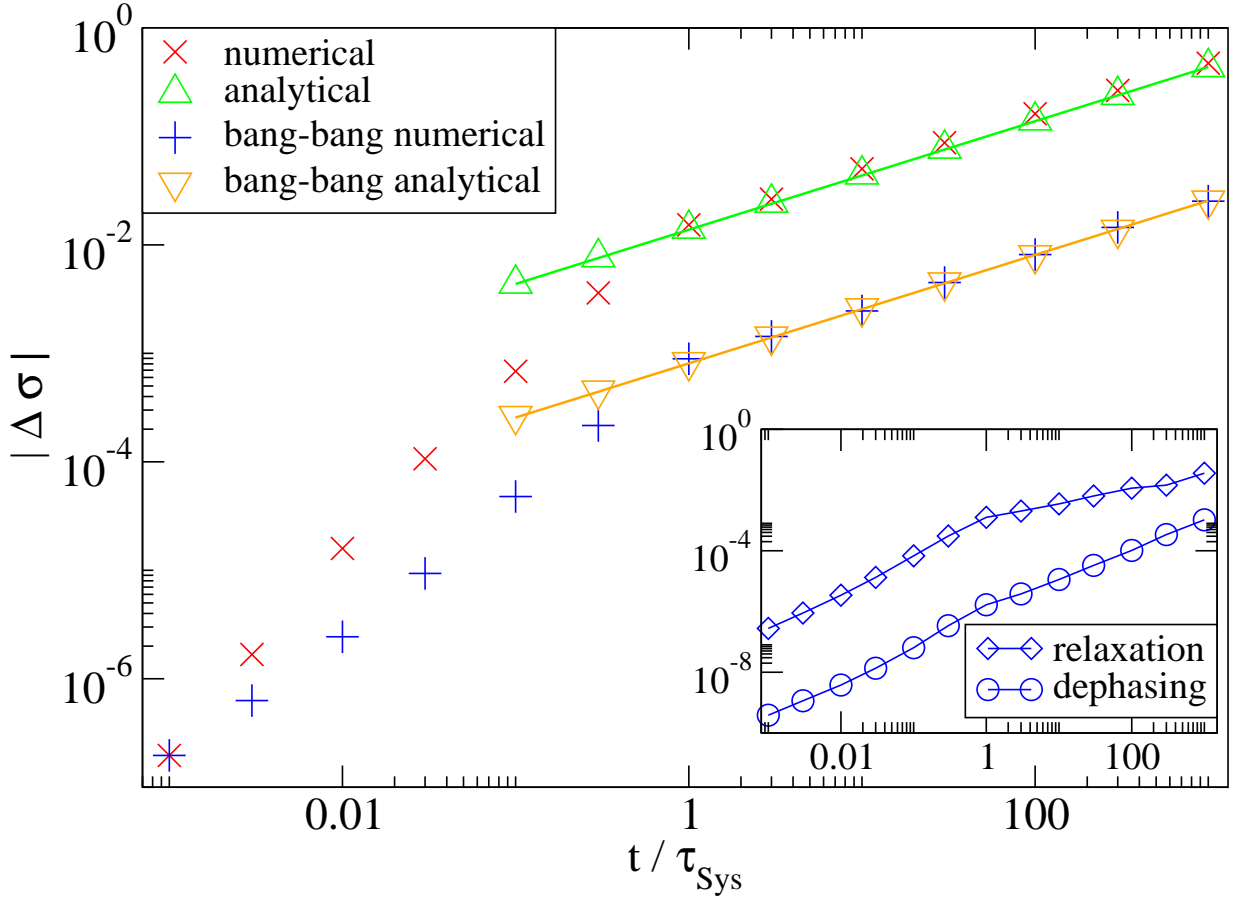}
 \caption{\label{devevolution}Time evolution of the rms deviations
 for bfl-induced random walks with and without bang-bang control at
 a coupling constant $\alpha=0.1$ and a typical flipping time scale
 $\tau_{\rm bfl}=10^{-2} \tau_{\rm Sys}$. The separation between
 two bang-bang pulses is $\tau_{\rm bb}=10^{-3}\tau_{\rm Sys}$. The
 straight lines are square-root fits of the analytical derived
 random walk model variances (plotted as triangles). Inset:
 Components of the deviations from the free precession trajectory
 that are parallel to it (dephasing) and perpendicular to it
 (relaxation/excitation) with bang-bang control.}
 \end{figure}

 We have numerically integrated Equ.~(\ref{ssesol}) and averaged
 the deviations of the random walk evolution from the unperturbed
 trajectories for times up to $100 \tau_{\rm Sys}$ over $N=10^3$
 realizations. Larger simulations have proven that convergence is
 already sufficient at this stage. We shall examine the
 root-mean-square (rms) deviations of this ensemble at given time
 points
 \begin{equation}\label{sigmarms}
 \Delta \vec{\sigma}_{\rm rms}(t) =  \sqrt{ \frac{1}{N}
 \sum_{j=1}^N \left( \vec{\sigma}^{\rm q}_j(t)-\vec{\sigma}^{\rm
 q}_{{\rm noisy},j}(t) \right)^2 }
 \end{equation}
 with and without bang-bang control. In other approaches, such as
 those based on master equations, one separates dephasing and
 relaxation.  Both are contained here in Equ.~(\ref{sigmarms}). We
 shall point out notable differences between these two channels.
 The deviation as a function of time is plotted in
 Fig.~\ref{devevolution}.

 The total deviations on intermediate time scales are suppressed by
 a ratio on the order of $10$. A detailed numerical analysis shows
 that {\em without} bang-bang suppression, the deviations parallel
 to the free precession trajectory (which correspond to dephasing)
 are of similar size to those perpendicular to free precession
 (which correspond to relaxation/excitation). In contrast, with
 bang-bang control, dephasing is almost totally absent as one can
 see in the inset of Fig.~\ref{devevolution}.

 The main double-logarithmic plot of Fig.~3 shows that on short
 time scales ($t \alt 0.1 \tau_{\rm Sys}$, which corresponds to
 $\alt 10$ random walk steps), deviations increase almost linearly
 in time. It is not until times on the order of $\tau_{\rm Sys}$
 that the noise-induced deviations start to behave as typical
 classical random walks, increasing as a square-root in time.

 \subsection{Analytical random walk models}

 We now develop analytical random walk models for our system. The
 random walk on the Bloch sphere is in general two-dimensional,
 consisting of both parallel and perpendicular deviations to the
 free evolution trajectory. Bang-bang control, as was seen in the
 above numerical results and as will be seen in the following
 analytical results, essentially reduces the random walk to
 one-dimension as only the perpendicular deviations remain
 significant. In the following, we restrict ourselves to the
 long-time (many random walk steps) regime.

 We first calculate for both cases the probability distributions of
 the deviations after one bfl flip (``one-step deviations'' in
 terms of the discrete random walk). 
The fluctuation of the period between $\tau_{\rm per}^{\pm}$
leads to dephasing, which can be evaluated at 
$\alpha\ll \epsilon_{\rm q},\Delta_{\rm q}$ to
 \begin{equation}
 \Delta\vec{\sigma}^{\rm bfl}_{\rm deph}  =  2 \pi \cos{\phi} 
\left(\frac{1}{\tau_{\rm per}^\pm} -  \frac{1}{\tau_{\rm per}}\right)
\tau_{\rm bfl}   
\simeq  \pm 2 \frac{\Delta_{\rm q}\epsilon_{\rm q}}{\Delta_{\rm q}^2+
\epsilon_{\rm q}^2} \alpha \tau_{\rm bfl}
 \end{equation}
where the prefactor $\cos{\phi}=\frac{\Delta_{\rm q}}
{\sqrt{\Delta_{\rm q}^2+\epsilon_{\rm q}^2}}$ takes the effective 
trajectory radius into account.
 
 For the relaxation/excitation effect of the noise, one has to use
the projection of the perturbation orthogonal to the free axis, using
$\sin{\eta} = \frac{\alpha \Delta_{\rm q}}{\epsilon_{\rm q}^2+\Delta_{\rm q}^2}$.
Furthermore this type of deviation also depends on the actual position 
of the spin on the Bloch sphere, e.g.\ there is no relaxation when the 
state is at one of the poles. 

\begin{figure}
 \includegraphics[width=0.9\columnwidth]{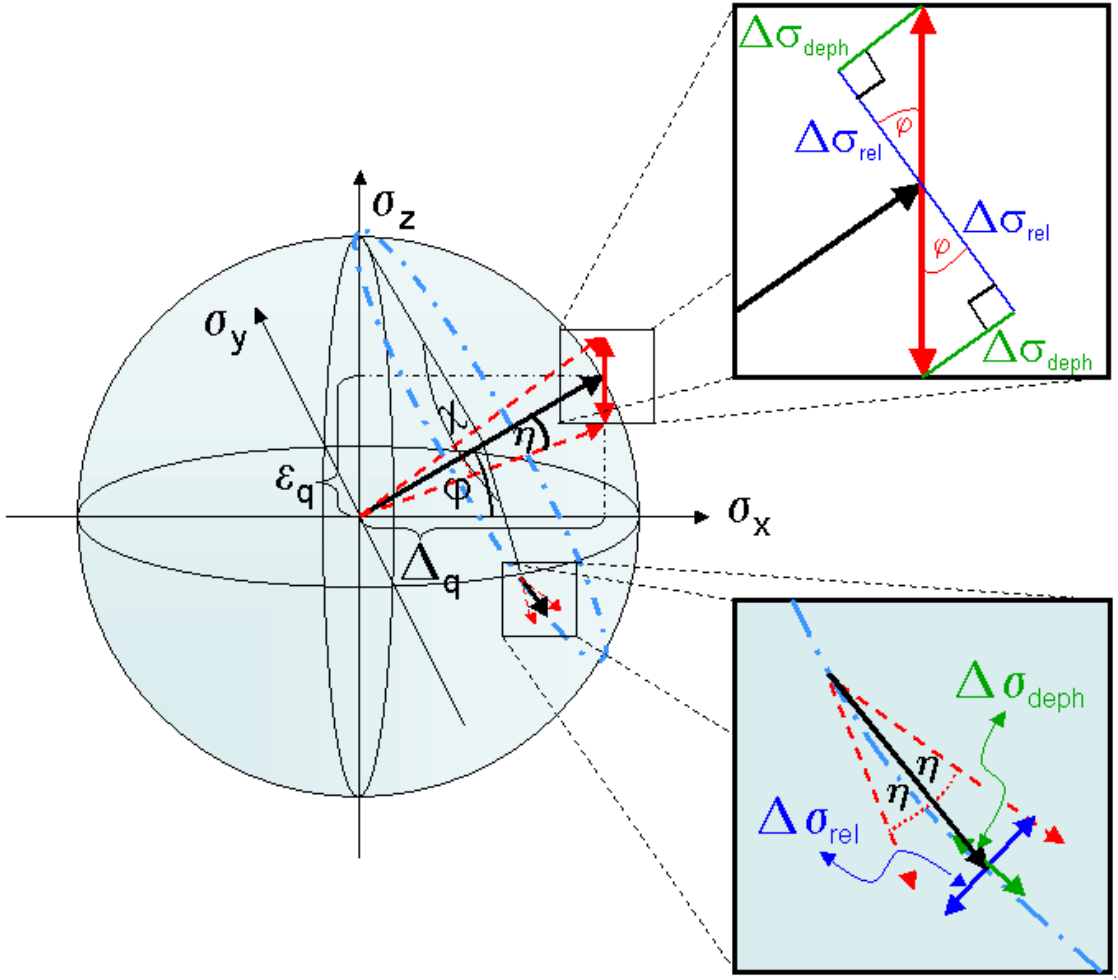}
 \caption{\label{bfl1rwstep} Plot of a typical one-step deviation
 from the unperturbed qubit trajectory with generic values for
 $\epsilon_{\rm q}$ and $\Delta_{\rm q}$. The fractions of the bfl
 fluctuations in $\hat{\sigma}_{\rm z}$-direction have to be
 distinguished with respect to their effects on the qubit: those
 that yield dephasing deviations that are parallel to the free
 precession trajectory (proportional to $\sin{\phi}$) versus 
relaxation/excitation deviations that are perpendicular (proportional
to $\sin{\eta}$). Both parts are additionally domineered by a factor 
of $\cos{\phi}$ due to the diminished radius of the trajectory 
starting from the initial state $\sigma_z=+1$. The impact of the 
relaxation/excitation generating part is furthermore depending on 
$\cos{\phi}$ as well as $\sin{\chi}$, the azimuth angle of the qubits 
present position.}
 \end{figure}

Averaging in rms-fashion over a full azimuthal cycle leads to a factor
of $1/\sqrt{2}$. Moreover, the impact of relaxation/excitation is
scaled down by an additional factor of $\cos{\phi}=
\frac{\Delta_{\rm q}}{\sqrt{\Delta_{\rm q}^2+\epsilon_{\rm q}^2}}$
corresponding to the projection of the Bloch vector onto 
the precession axis, which furthermore decreases the deviation
angle.
In total, using $\tau_{\rm per}^\pm \simeq \tau_{\rm per}$ to first 
order in $\alpha$, we find
 \begin{eqnarray}
   \Delta\vec{\sigma}^{\rm bfl}_{\rm rel} =  2 \pi \cos{\phi} 
\sin{\eta} \frac{1}{\sqrt{2}} \cos{\phi} \frac{\tau_{\rm bfl}}
{\tau_{\rm per}^\pm} \simeq \sqrt{2} \frac{\Delta_{\rm q}^3}
{(\epsilon_{\rm q}^2+\Delta_{\rm q}^2)^{3/2}} \alpha \tau_{\rm bfl}.
 \end{eqnarray}
 Our root-mean-square measure of the impact of the noise, Equ.~(5),
 does not handle these two kinds of deviations separately, but
 rather adds them up to:
 \begin{eqnarray}\label{bfl1step}
 \Delta\vec{\sigma}^{\rm bfl}_{\rm total} \hspace{2mm} = \hspace{2mm}
\sqrt{{\Delta\vec{\sigma}^{\rm bfl}_{\rm deph}}^2+
{\Delta\vec{\sigma}^{\rm bfl}_{\rm rel}}^2}  & = & 
 \sqrt{4 \frac{\Delta_{\rm q}^2\epsilon_{\rm q}^2}{(\Delta_{\rm q}^2+
\epsilon_{\rm q}^2)^2} + 2\frac{\Delta_{\rm q}^6}
{(\epsilon_{\rm q}^2+\Delta_{\rm q}^2)^3}} \alpha \tau_{\rm bfl} \nonumber \\
 & = & \frac{1}{(\Delta_{\rm q}^2+\epsilon_{\rm q}^2)^{3/2}} 
\sqrt{4(\Delta_{\rm q}^2 + \epsilon_{\rm q}^2) \Delta_{\rm q}^2
 \epsilon_{\rm q}^2 +2\Delta_{\rm q}^6}
\alpha \tau_{\rm bfl} 
 \end{eqnarray}

Our rms treatment disregard the different types of decoherence, dephasing 
and relaxation/excitation, corresponding to phase and bit-flip errors
respectively. This is no crucial drawback but merely lies
in the nature of our generic situation. If needed, both components can
be isolated.  

\begin{figure}[h]
 \includegraphics[width=0.8\columnwidth]{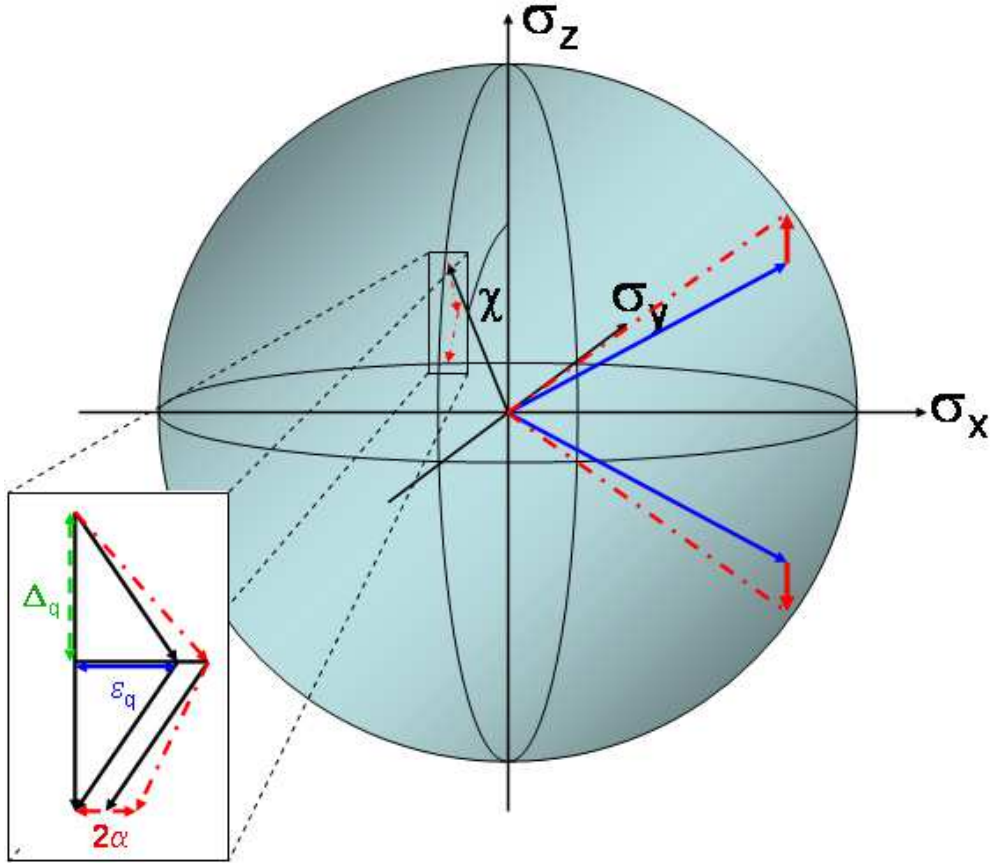}
 \caption{\label{bb1rwstep} Sketch of a maximal one-step deviation during
 a bang-bang modulated cycle, which appears if the bfl state flips precisely
 at the intermediate bang-bang pulse time. The dephasing part of deviation
 evidently averages out, while a relaxating aberrance arise proportional
 to the noise-coupling constant $\alpha$.}
 \end{figure}

 The derivation of the maximal one-step deviation for the bang-bang
 controlled situation has to be handled differently. The deviation
 resulting from a bfl flip during a bang-bang pulse period is
 maximal if the step happens exactly at the moment of the second
 qubit spin-flip (\textit{i.e.}, in the middle of the bang-bang
 cycle).  When this happens, the refocusing evolution has in its
 first half a drift, for example, to the ``right'' (compare to
 Fig.~\ref{bb1rwstep}) and in the last half an equal
 aberration.\\

 The resulting one-step deviation appears to be on the order of $2
 \alpha \tau_{\rm bb}$.  However, this is scaled down by a factor
 of $1/\sqrt{2}$, as the impact of the aberration in $x$-direction
 is proportional to a factor of $\sin\chi$, where $\chi$ denotes
 the longitudinal angle of the present spin position on the Bloch
 sphere (see Fig.~\ref{bb1rwstep}). This is because the
 $\sigma_z$-component of the noisy evolution does not influence the
 qubit, if it is near the $\sigma_z = \pm 1$-state and its
 influence is suppressed correspondingly in between. As we are
 mainly interested in mean aberrations after many random walk
 steps, we simply average the maximal one-step deviation over one
 precession period in the usual rms manner to obtain
 \begin{eqnarray}\label{bb1step}
 \left<\Delta\vec{\sigma}^{\rm bb}_{\rm max}\right> & = &
 \sqrt{\frac{1}{2 \pi}\int_0^{2\pi} \sin^2\chi 4 \alpha^2 \tau_{\rm bb}^2 \,d\chi} \nonumber \\
 & = & \sqrt{2}\alpha \tau_{\rm bb} .
 \end{eqnarray}
 Obviously, this variance only contributes to relaxation.

 In the long time limit, we replace the fluctuating number of
 random walk steps for a given time $\Delta t$ of noisy evolution
 by its expectation value $N_{\rm bfl}=\Delta t/\tau_{\rm bfl}$. 
This allows us to use the number of random walk steps as time parameter
\cite{WeissRW}. This simplification does not introduce
 significant error, as the relative number variation for $\Delta t$
 scales as $\frac{\sqrt{\Delta t/\tau_{\rm bfl}}} {\Delta
 t/\tau_{\rm bfl}}=\sqrt{\tau_{\rm bfl}/\Delta t} \rightarrow 0$ in
 our preferred long-time limit. We encounter two different
 one-step-distributions, depending on whether the numeration of the
 step is an odd or even (corresponding to an ``up'' or ``down''
 state of the bfl). For definiteness, we assume the bfl is
 initially in its ``upper'' state, which is of no influence on the
 long time limit as the memory to the initial state is already
 erased. The step-size distribution of the bfl model in our small
 deviation regime is given from Poisson statistics
 \begin{equation}
 \Phi_{\rm odd/even}^{\rm bfl} (x)  =  {e^{\mp x / \beta} \theta(\pm x)
 \over \beta}
 \end{equation}
 with $\beta = \frac{\sqrt{5}}{2} \alpha \tau_{\rm bfl}$ the typical one-step
 deviation as calculated in Equ.~\ref{bfl1step}. $\theta(x)$ denotes
 the Heaviside step function. We neglect the correlations between
 transverse and perpendicular deviations as we expect them to
 average out in the long-time limit.

 For the bang-bang suppressed random walk, the flipping positions
 of the bfl-noise sign in the bang-bang time-slots are essentially
 randomly distributed as long as $\tau_{\rm bb}\ll\tau_{\rm bfl}$.
 That is why we find a constant step-size distribution between zero
 and a maximum divergence of $\gamma = {2 \alpha \tau_{\rm bb}\over
 \sqrt{2}}$ (see Equ.~\ref{bb1step}), namely
 \begin{eqnarray}
 \Phi_{\rm odd/even}^{\rm bb} (x) & = & {\theta(\pm x) \theta\left(
 \gamma \mp x \right) \over \gamma}.
 \end{eqnarray}

 By means of these one-step probability distributions, we are able
 to calculate via the convolution theorem the distributions for
 $2N_{\rm bfl}$-step random walks. Specifically, they are the
 inverse Fourier transforms of the $N_{\rm bfl}$-fold products of
 the Fourier transforms of the two-step distribution
 \cite{WeissRW}. For the case without bang-bang control, we find
 \begin{eqnarray}\label{phibfl}
 \Phi_{\rm 2N_{\rm bfl}}^{\rm bfl}(x) & = & \mathscr{F}^{-1} \left[ \left(
 \mathscr{F}
 \left[ \Phi_2^{\rm bfl}\right]\right)^{N_{\rm bfl}} \right]  \nonumber \\
 & = & \int_{-\pi}^{\pi}
 \frac{dk}{2\pi\beta^{2N_{\rm bfl}}}  e^{-i k x}
 \left(\frac{1}{1-2\cos(k) e^{-1/\beta} + e^{-2/\beta}} \right)^{N_{\rm bfl}}
 \end{eqnarray}
 whereas for the case with bang-bang control, it is
 \begin{eqnarray}\label{phibb}
 \Phi_{\rm 2N_{\rm bfl}}^{\rm bb}(x) & = & \mathscr{F}^{-1} \left[ \left(
 \mathscr{F}
 \left[ \Phi_2^{\rm bb}\right]\right)^{N_{\rm bfl}} \right]  \nonumber \\
 & = & \int_{-\pi}^{\pi} \frac{dk}{2\pi\gamma^{2N_{\rm bfl}}} e^{-ikx}
 \left(\frac{[1-\cos((\gamma+1)k)]}{[1-\cos(k)]} \right)^{N_{\rm bfl}}
 \end{eqnarray}
 with $\mathscr{F}$ and $\mathscr{F}^{-1}$ denoting the discrete
 Fourier transformation and its inverse, respectively.

 Already for random walk step-numbers on the order of 10, the
 resulting distributions are almost Gaussian. Their standard
 deviations give the rms deviations of the random walk models
 plotted in Fig.~\ref{devevolution}. As in the numerical
 simulations at long times, they grow as a square-root of the
 number of steps. As one can recognize, the underlying two-step
 distributions in the $k$-space (\textit{i.e.}, the functions in
 the large brackets of Equs.~(\ref{phibfl}) and (\ref{phibb})) are
 symmetric and differentiable around zero such that the above
 integrals can be evaluated analytically using the saddle point
 approximation (the small parameter is $k$, which is justified at
 least qualitatively in our bounded variable integral). We find for
 their variances in real space representation
 \begin{equation}
 \label{sigmabfl} \Delta\sigma_{\rm bfl}(N_{\rm bfl}) = \sqrt{N_{\rm bfl}} \beta
 = \sqrt{N_{\rm bfl}} \frac{\sqrt{5}}{2} \alpha \tau_{\rm bfl}
 \end{equation}
 for the case without bang-bang control and
 \begin{equation}
 \label{sigmabb} \Delta\sigma_{\rm bb}(N_{\rm bfl}) = \frac{\sqrt{N_{\rm
 bfl}}}{2} \gamma = \sqrt{\frac{N_{\rm bfl}}{2}}\alpha\tau_{\rm bb}
 \end{equation}
 for the case with it. In the large-$N_{\rm bfl}$ limit, this model
 shows excellent agreement with the numerical simulations.

\begin{figure}[h]
 \includegraphics[width=0.4\columnwidth]{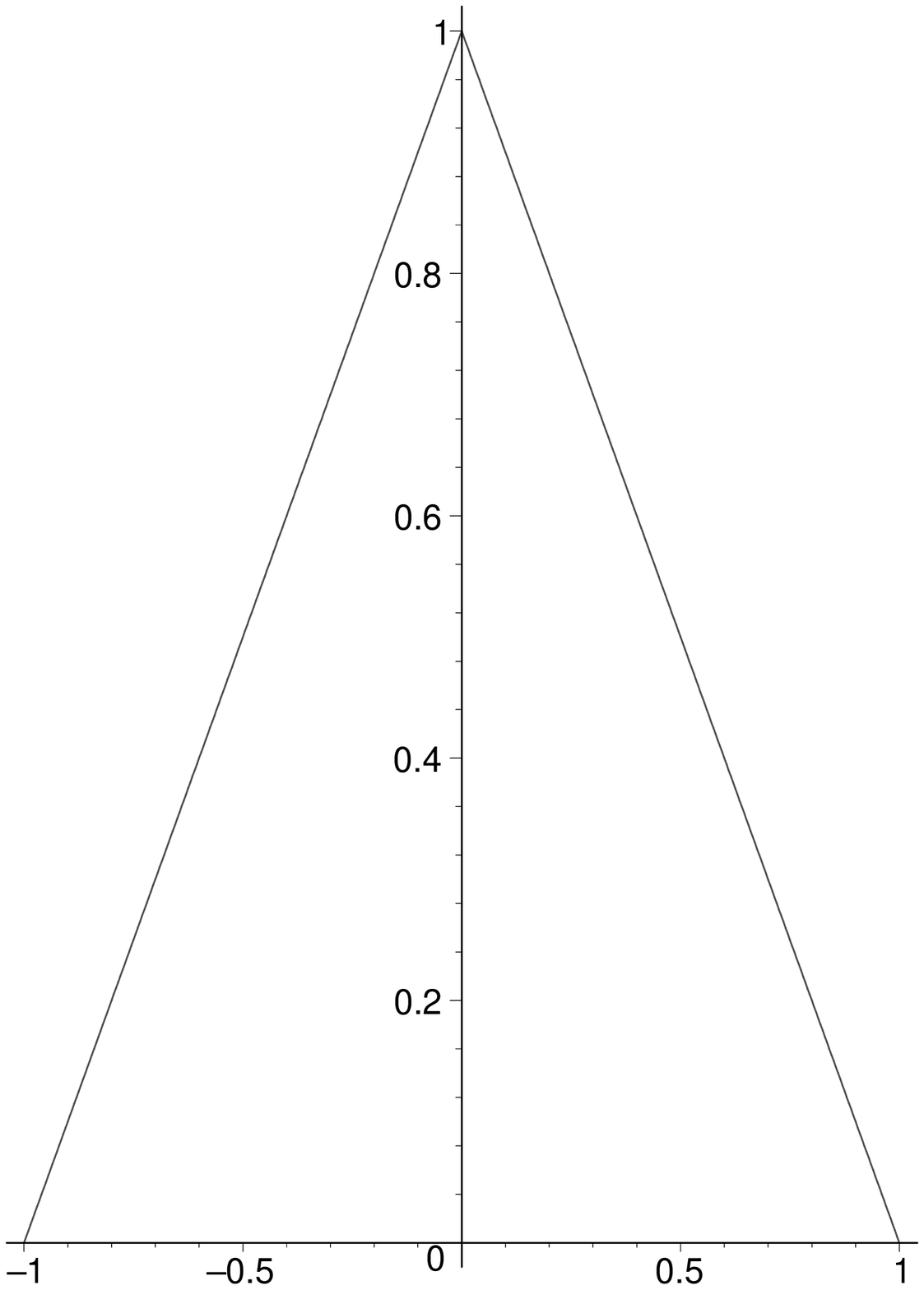}
 \includegraphics[width=0.4\columnwidth]{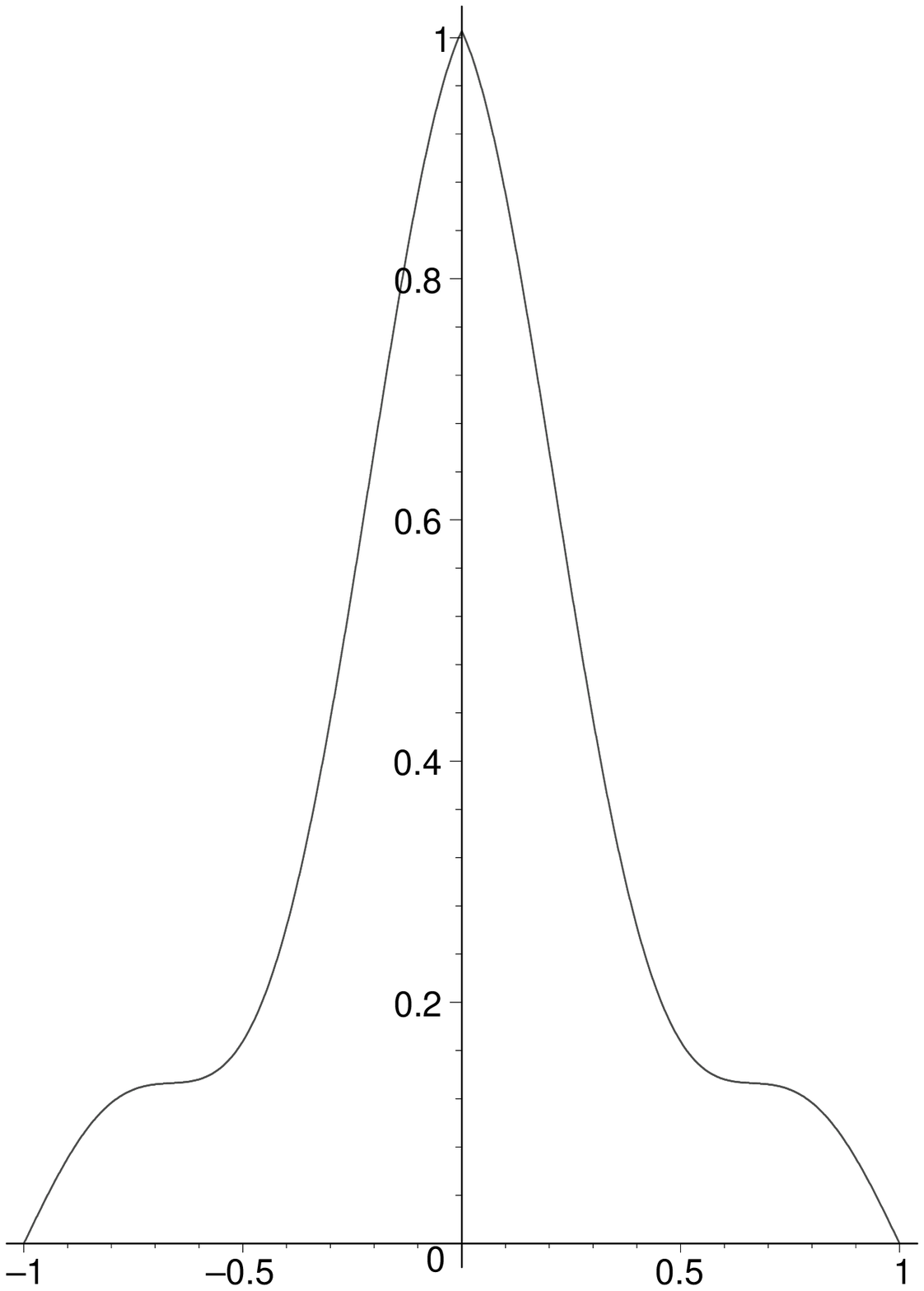}
 \caption{\label{2stepdistr} Comparison of two-step distributions
 for the random walks with bang-bang control when the bang-bang
 pulses are taken to be $\delta$-functions (left) versus a
 continuous sine wave $\sin(\frac{\pi}{\tau_{bb}}t)$ (right).  For
 clarity, the $y$-axis is rescaled to the maximum values of the
 distributions, while the $x$-axis is given in units of $\gamma$.}
 \end{figure}

 At first sight, treating bang-bang pulses as $\delta$-function
 impulses appears to be an extraordinarily strong assumption,
 especially because in a physical implementation, the large
 bandwidth associated with very short pulses could excite other
 noise sources. However, this $\delta$-function impulse
 approximation is only for technical simplification. In fact, going
 to the other extreme of a wide, continuous pulse of the form
 $\sin(\frac{\pi}{\tau_{bb}}t)$ would also refocus our bfl-noise
 over the course of its periods. Comparing the two-step deviation
 distributions arising from $\delta$-function impulses versus
 continuous sine waves, one obtains for the $\delta$-function case
 \begin{eqnarray}
 \Phi_2^{\rm inf}(x) & = & \frac{|\gamma-x|}{\gamma} \theta(\gamma-x)
 \theta(\gamma+x)
 \end{eqnarray}
 and for the continuous sine wave case
 \begin{eqnarray}
 \Phi_2^{\rm cont}(x) & = & \left\{ \left[ \frac{\pi}{2 \gamma}
 +\frac{\pi}{4 \gamma} \cos \left( 2\pi \frac{x}{\gamma} \right)
 \right] \left( 1 - \frac{x}{\gamma} \right) + \frac{3}{16 \gamma}
 \sin \left( 2\pi \frac{x}{\gamma} \right) \right\}
 \theta(\gamma-x) \theta(\gamma+x)
 \end{eqnarray}
 These distributions are depicted in Fig.~\ref{2stepdistr}.  One
 recognizes that in fact the distribution arising in the continuous
 sine wave case is narrower (and therefore indicates \textit{more}
 effective noise suppression) than the $\delta$-function impulse
 case, with the drawback of leaving less free evolution time for
coherent operation.

 \subsection{Distributions of the random walks deviation}

\begin{figure}[h]
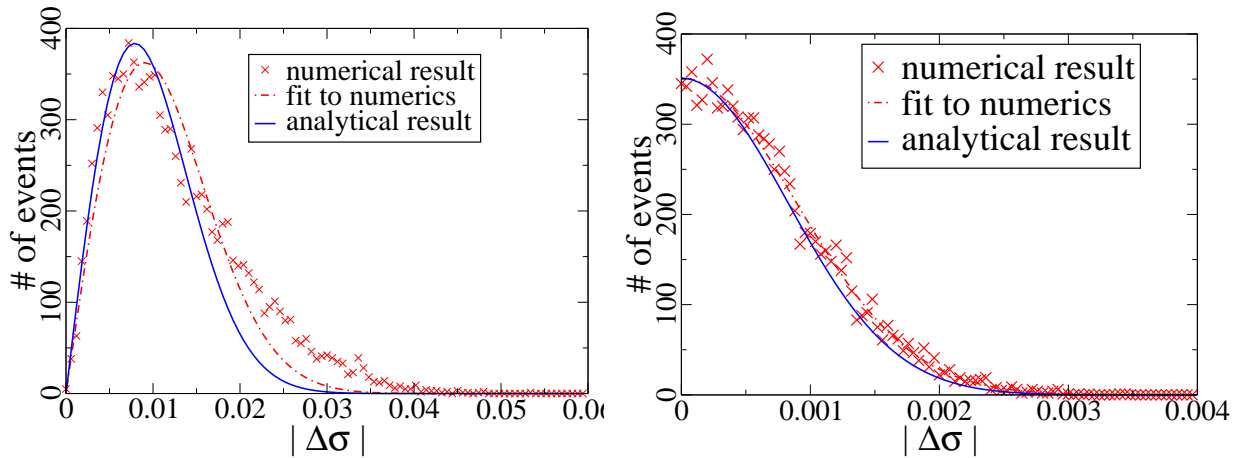

 \includegraphics[width=0.49\columnwidth]{bfl_hist.eps}
 \includegraphics[width=0.49\columnwidth]{bb_hist.eps}
 \caption{\label{distributions}Histograms of the deviation from
 free evolution both without bang-bang control (left) and with
 bang-bang control (right).   Also plotted are fits to the expected
 two- and one-dimensional random walk statistics respectively
 associated with the uncontrolled and controlled cases. Numerical
 data were collected over $10^4$ realizations at a fixed time
 $t_0=\tau_{\rm Sys}$ defined such that $\tau_{\rm bfl}= 0.01
 \tau_{\rm Sys}$ and thus $N_{\rm bfl}=\tau_{\rm Sys}/\tau_{\rm
 bfl}=100$ steps. (NB: The $x$-axis scale of the right graph
 depicting the bang-bang controlled case is 15 times smaller than
 that of the left graph depicting the uncontrolled case.)}
 \end{figure}

 Beyond predicting the variances of the random walks, our analysis
 also allows evaluation of their full probability distributions. We
 compare them to numerics with and without bang-bang compensation
 by use of simulations with $10^4$ realizations at an evolution
 time $t_0=\tau_{\rm Sys}$. The numerical histograms of the
 deviations with their respective one-
 and two-dimensional Gaussian fits are shown in Fig.~\ref{distributions}.\\

 We observe that not only the distribution obtained with bang-bang
 control is much narrower than the distribution obtained without
 it, but also that its shape is qualitatively different.  The
 maximum of the bang-bang controlled distribution is at zero error.
 In contrast, the uncontrolled distribution has its maximum at a
 finite error $|\Delta \sigma|_{\rm max} \approx 0.01$, and it has zero
 probability of zero error. This reflects the one-dimensional
 nature of the bang-bang controlled random walk in contrast to the
 two-dimensional nature of the uncontrolled random walk.

 \subsection{Bang-bang control working as a high-pass filter}

 In order to measure the degree of noise suppression due to
 bang-bang control, we define the suppression factor
 $\mathcal{S}_{t_0}$ as follows for a given evolution time $t_0$
 \begin{eqnarray}
 \mathcal{S}_{t_0}(\tau_{\rm bfl}/\tau_{\rm bb}) & \equiv &
 \frac{\Delta\vec{\sigma}_{\rm rms}^{\rm
 bfl}(t_0)}{\Delta\vec{\sigma}_{\rm rms}^{\rm bb}(t_0)}.
 \end{eqnarray}

\begin{figure}[h]
 \includegraphics[width=0.99\columnwidth]{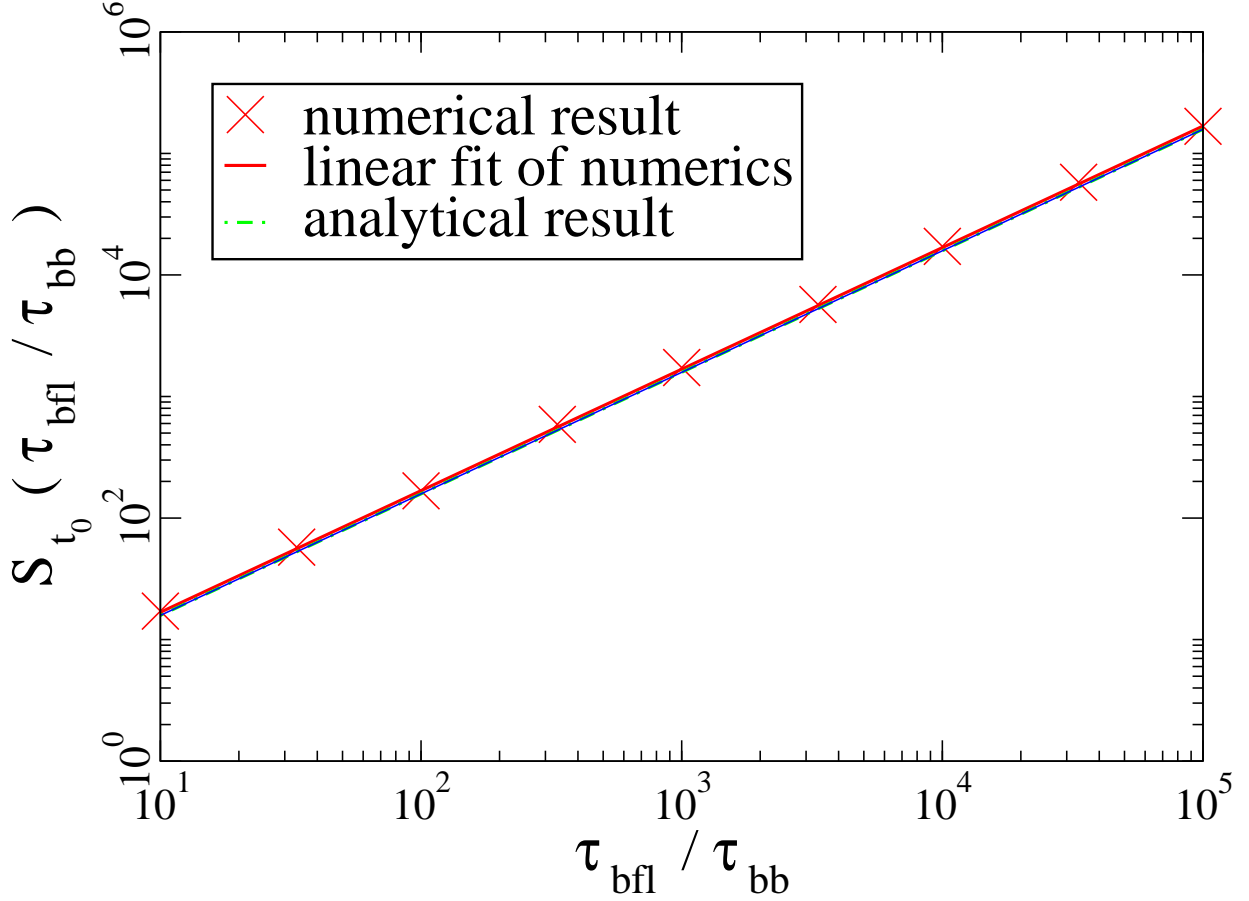}
 \caption{\label{supfactor}The suppression factor
 $\mathcal{S}_{t_0} (\tau_{\rm bfl}/\tau_{\rm bb}) =
 \frac{\Delta\vec{\sigma}_{\rm rms}^ {\rm
 bfl}(t_0)}{\Delta\vec{\sigma}_{\rm rms}^{\rm bb}(t_0)}$ evaluated
 for $t_0=\tau_{\rm Sys}$ as a function of the ratio of the mean
 switching time $\tau_{\rm bfl}$ and the bang-bang pulse separation
 $\tau_{\rm bb}$.}
 \end{figure}

 We now systematically study the dependence of $\mathcal{S}_{t_0}$
 on $\tau_{\rm bfl}/\tau_{\rm bb}$ for a constant mean bfl
 switching rate $\tau_{\rm bfl}=10^{-2}\tau_{\rm sys}$ at a fixed
 evolution time $t_0=\tau_{\rm sys}$. The numerical data in
 Fig.~\ref{supfactor} show that the suppression efficiency is
 linear in the bang-bang repetition rate, $S_{\tau_{\rm sys}}=\mu
 \tau_{\rm bfl}/\tau_{\rm bb}$. The numerically derived value of
 the coefficient, $\mu_{\rm numerical} \approx 1.679$, is in
 excellent agreement with the analytical result $\mu_{\rm
 analytical}=\sqrt{5/2} \simeq 1.581$ from our saddle point
 approximation, Equs.~(\ref{sigmabfl}) and (\ref{sigmabb}).\\
~\\

 This small discrepancy between the numerical and analytical
 results is due to the fact that the analytical calculations
 neglect correlations between the parallel and perpendicular
 components of the random walk. This leads to an underestimate of
 the rms-deviation $\Delta\vec{\sigma}_{\rm rms}^{\rm bfl}$ in the
 case without bang-bang control (compare also to
 Fig.~\ref{distributions}). Therefore, we have quantitatively
 proved our qualitative intuition: bang-bang control affects the
 bfl noise signal like a high-pass filter, an effect that one of
 the authors has generally predicted for dynamical decoupling
 techniques \cite{LVK99a}.

 \section{Limitations due to pulse inaccuracies}

 Thus far, we have tacitly assumed that one could apply perfect,
 zero-width $\pi$-pulses along exactly the
 $\hat{\sigma}_{\rm x}$-axis of the Bloch sphere.  We now take into
 account that the control pulses themselves typically will have
 slight fluctuations in their duration or polarization that
 interfere with the desired refocusing. As already shown at the end
 of Section IV.B, the restriction of pulses to infinitesimal
 duration can be significantly relaxed.  We now investigate to what
 extent the restriction to perfect pulses can be relaxed.

 \subsection{Two generic types of bang-bang inaccuracies}

 We essentially analyze two generic types of errors that could
 occur in the control apparatus when trying to apply $\pi$-pulses
 in $\hat{\sigma}_{\rm x}$-direction.  One, the duration of each
 pulse could exhibit fluctuations, resulting in fluctuations in the
 rotation-angle around the desired value of $\pi$. Two, the
 polarization axis could suffer from directional deviations around
 the desired value of $\hat{\sigma}_{\rm x}$. Assuming the
 statistical independence of each pulse error, we expect for both 
types of imperfections a random-walk-like behavior of increasing
 deviations compared to evolutions with perfect pulses.

 \subsubsection{One-dimensional pulse error (dephasing)}

\begin{figure}[h]
 \includegraphics[width=0.99\columnwidth]{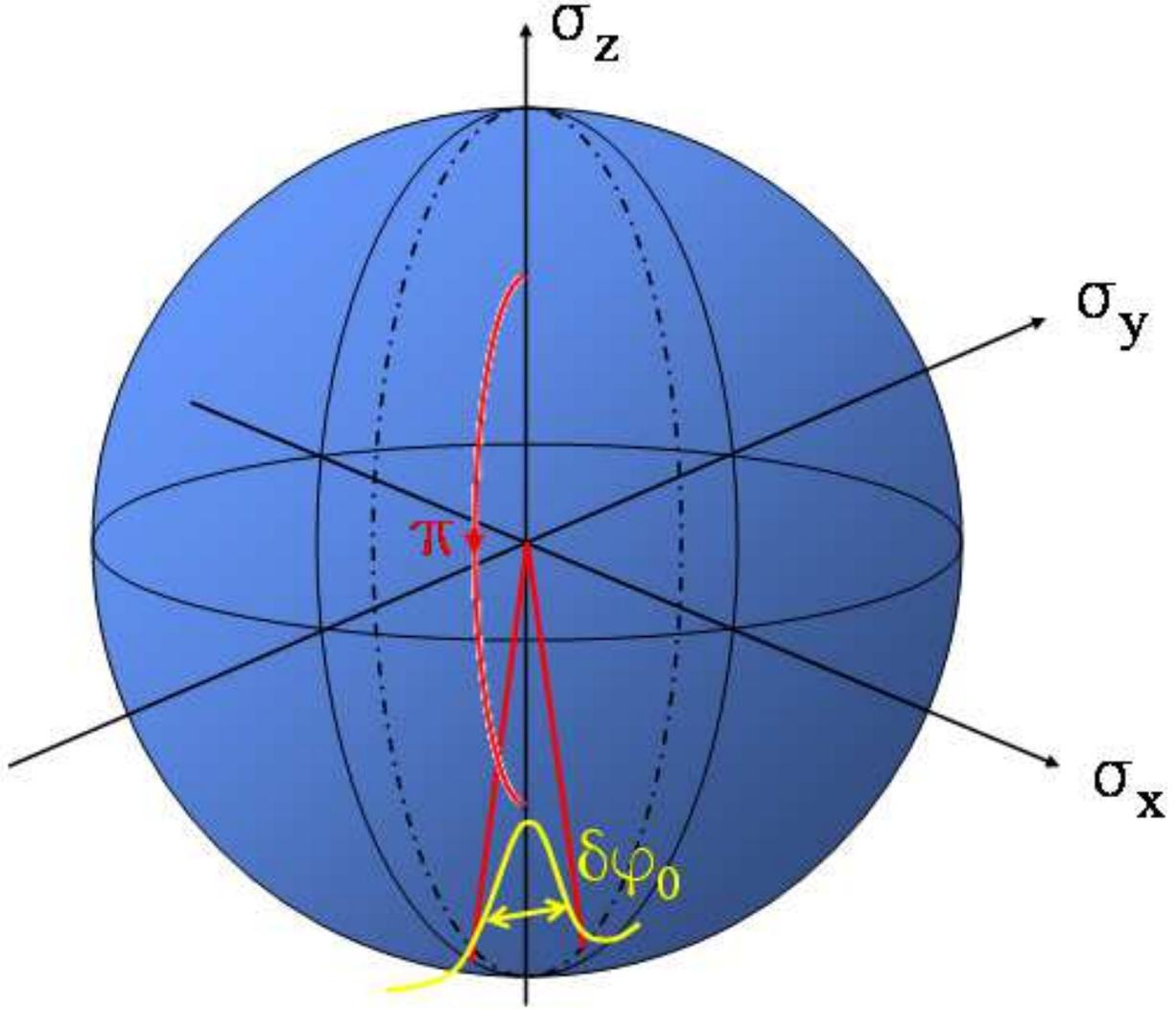}
 \caption{\label{bberror1}Sketch of one-dimensional bang-bang
 aberration. The variations $\delta\phi_0$ of the rotation angle
 around the desired value of $\pi$ leads to slight deviations in
 parallel to the permitted dynamical direction, thus generating
 dephasing.}
 \end{figure}

 We make the quite general assumption that we may model the
 one-dimensional phase fluctuation of the imperfect bang-bang
 pulses $\phi_j(x)$ as a Gaussian distribution of the pulse
 durations and therefore of the rotation angles around their
 intended value $\pi$.  This assumption should be valid for many
 physical situations, \textit{e.g.}, if the inaccuracy is due to
 electromagnetic noise in the pulse generator.  The Gaussian is
 parameterized by its standard deviation $\delta\phi_0$ (see
 Fig.~\ref{bberror1}). Thus, the corresponding pulse angle
 aberration of the $j$th step is given by
 \begin{eqnarray}
 \label{1deq}
 \phi^{\rm 1d}_j(x) & = & \frac{1}{\sqrt{2\pi}\delta\phi_0}
 e^{-\frac{x^2} {2 \delta\phi_0^2}} .
 \end{eqnarray}
 Having assumed a Gaussian distribution, we can exactly evaluate
 the distributions of the $N$-step deviation $\Delta\Phi_N$ (which
 are usually given as $N$-fold time-convoluted integrals) as
 follows by use of the convolution theorem
 \begin{eqnarray}
 \label{1dDPhi}
 \Phi^{\rm 1d}_N & = & \mathscr{F}^{-1} \left[ \Pi_{j=1}^N
 \tilde{\phi}_j \right] \nonumber \\
 & = &  \frac{1}{\sqrt{2\pi N} \delta\phi_0} e^{-\frac{x^2}{2 N
 \delta\phi_0^2}}
 \end{eqnarray}
 with $\tilde{\phi}^{\rm 1d}_j=\mathscr{F}[\phi_j]$ denoting the
 Fourier transform of $\phi^{\rm 1d}_j$ and $\mathscr{F}^{-1}$
 denoting the inverse Fourier transform.

 Therefore the rms displacement in the random walk increases as a
 square-root in the number $N$ of bang-bang pulses:
 $\delta\phi_N=\sqrt{N}\delta\phi_0$. Equivalently, the dephasing
 grows as square-root in time
 \begin{eqnarray}
 \label{1dbbtime}
 \delta\phi(t)=\sqrt{t/\tau_{\rm bb}}\delta\phi_0
 \end{eqnarray}
 on the time scale of our coarse-graining (which is here given as
 $\tau_{\rm bb}$).

 \subsubsection{Two-dimensional pulse error: dephasing and
 relaxation/excitation}

\begin{figure}[h]
 \includegraphics[width=0.99\columnwidth]{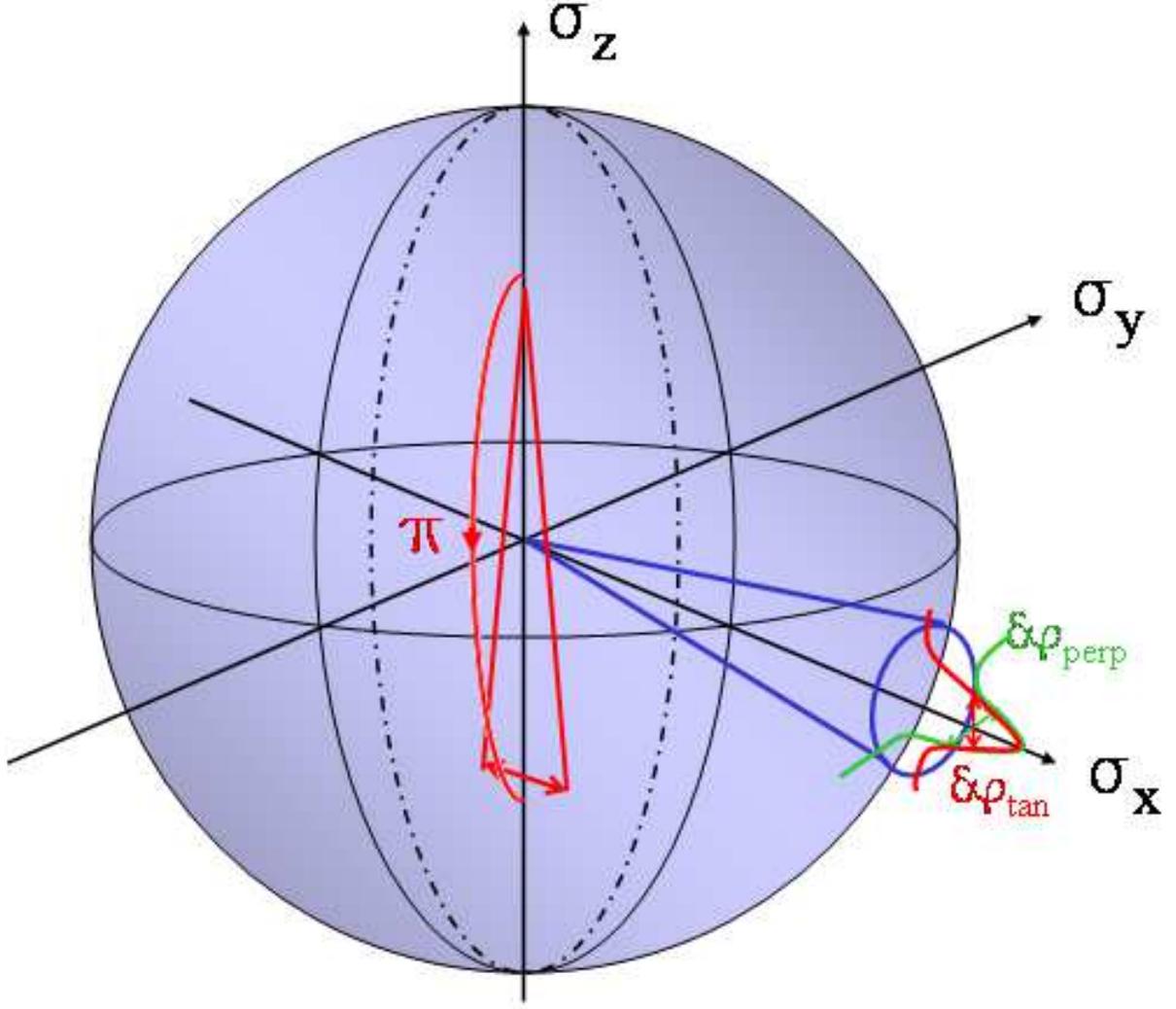}
 \caption{\label{bberror2}Sketch of two-dimensional bang-bang
 aberration. To first order, variations $\delta\phi_{\rm perp}$ of the
 rotation axis perpendicular to the connection vector between
 $\sigma_{\rm x}$ and the qubit state (here for simplicity:
 $\sigma_{\rm z}=+1$) do not influence the intended spin-flip,
 whereas the variations $\delta\phi_{\rm tan}$ along this line causes 
 deviations on the Bloch sphere perpendicular to the permitted evolution 
 trajectories (therefore producing relaxation or excitation).}
 \end{figure}

 A similar argument works when there are also fluctuations around
 the desired $\hat{\sigma}_{\rm x}$ rotation axis. Each individual
 variation of the axis can be split into two components: (1)
 $\delta\phi_{\rm perp}$, which is perpendicular to the connecting
 vector between the $\hat{\sigma}_{\rm x}$-axis and the qubit state
 $\vec{\sigma}(t)$ on the Bloch sphere, and (2) $\delta\phi_{\rm
 tan}$, which is transverse to it (see Fig.~\ref{bberror2}).  To
 first order, the perpendicular part does not disturb the intended
 spin-flip \footnote{The attentive reader might object that a
 spin-flip around a different axis on the xy-equator doesn't
 commute with pure $\hat{\sigma}_{\rm x}$-dynamics, but rather with
 something nearby. As we do not consider any $\hat{\sigma}_{\rm
 y}$-components yet, we do not bother about the minimal distortion
 of pure $\hat{\sigma}_{\rm x}$-manipulations, which can be
 estimated to be on the $2^{\rm nd}$ order of the aberration parameter
 $\delta\phi_0$ that we assumed to be very small anyway.}. However,
 the transverse part does cause a deviation from the ideal
 spin-flip in a direction toward or away from the previous qubit
 state. (Therefore, it produces relaxation or excitation, as its
 effect is orthogonal to the free $\hat{\sigma}_{\rm
 x}$-evolution.) Consequently, in a statistical average we only
 have to consider $1/\sqrt{2}$ of the typical total mean
 $\delta\phi_0$ of the aberration. The effect of a $\pi$-rotation
 around an axis tilted by an angle $\delta\phi_{\rm tan}$ is a
 deviation $2\delta\phi_{\rm tan}$ from the trajectory of the
 perfect evolutions; thus we receive altogether a deviation on the
 order of $\sqrt{2}\delta\phi_0$.

 Therefore, we obtain analogously to Equ.~(\ref{1deq}) for each
 single step distribution
 \begin{eqnarray}
 \label{2deq}
 \phi^{\rm 2d}_j(x) & = & \frac{1}{\sqrt{2\pi} \sqrt{2} \delta\phi_0}
 e^{-\frac{x^2} {4 \delta\phi_0^2}} ;
 \end{eqnarray}
 and analogously to Equ. (\ref{1dDPhi}) for the deviation after $N$
 steps
 \begin{eqnarray}
 \label{2dDPhi} \Phi^{\rm 2d}_N & = &  \frac{1}{\sqrt{2\pi} \sqrt{2N}
 \delta\phi_0} e^{-\frac{x^2}{4 N \delta\phi_0^2}}.
 \end{eqnarray}
 Equivalently, in terms of the time $t$
 \begin{eqnarray}
 \label{2dbbtime}
 \delta\phi(t)=\sqrt{2 t/\tau_{\rm bb}} \cdot \delta\phi_0 .
 \end{eqnarray}

 \subsection{Numerical and analytical results}

 In the same manner as our previous integrations of a stochastic
 Schr\"odinger equation, we numerically simulate qubit dynamics
 under inaccurate pulses. In the first instance, we work without
 bfl-noise to verify our analytical random walk model. Later, we
 add the bfl-noise in order to study the competition between the
 two sources of error.

 \subsubsection{Random walk due to inaccurate bang-bang pulses only}

 We analyze deviations on the Bloch sphere between the noiseless
 case trajectories that occur when the bang-bang pulses are perfect
 and those when they are not. As per Equ.~(\ref{sigmarms}), we
 calculate the rms-deviation over ensembles of $N=10^3$
 realizations. As a representative time point, we once again choose
 $t_0=\tau_{\rm Sys}$.  This is because, as explained in the
 discussion surrounding Fig.~\ref{devevolution}, this time scale
 should exhibit neither short-time effects nor near-total
 decoherence. From Equs.~(\ref{1dbbtime}) and (\ref{2dbbtime}), it
 immediately follows that for the mean deviations at $t_0$ if there
 are phase errors
 \begin{eqnarray}
 \label{1danaleq}
 \Delta\sigma^{\rm 1d}_{\rm bb}(t_0) & = & \sqrt{N_{\rm bb}} \delta\phi_0
  \hspace{2mm} = \hspace{2mm} \sqrt{\frac{t_0}{\tau_{\rm bb}}}
  \delta\phi_0,
 \end{eqnarray}
 and if there are axis errors
 \begin{eqnarray}
 \label{2danaleq} \Delta\sigma^{\rm 2d}_{\rm bb}(t_0) & = & \sqrt{2 N_{\rm
 bb}} \delta\phi_0  \hspace{2mm} = \hspace{2mm} \sqrt{2
 \frac{t_0}{\tau_{\rm bb}}} \delta\phi_0.
 \end{eqnarray}
~\\
 \begin{figure}[h]
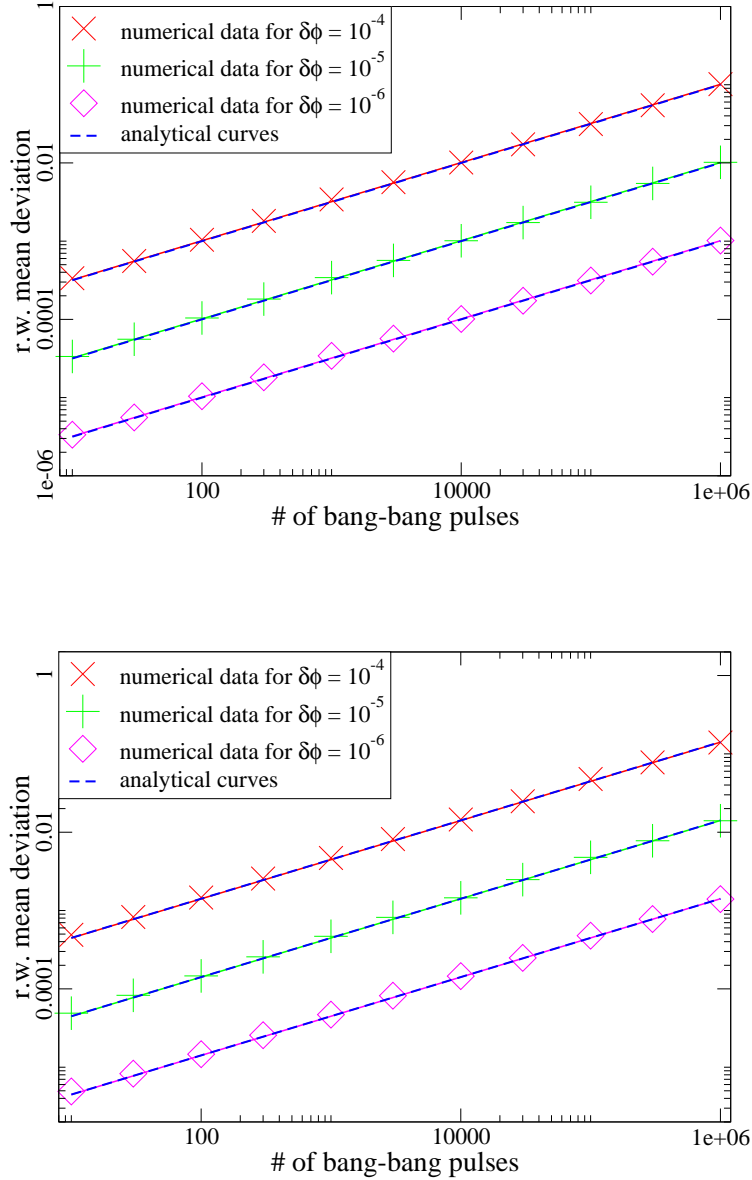

 \includegraphics[width=0.6\columnwidth]{bb1fehler.eps}\\
~~\\
~~\\
 \includegraphics[width=0.6\columnwidth]{bb2fehler.eps}
 \caption{\label{bbfehler}Plot of the one- respectively two-dimensional
 imperfectly bang-bang pulsed evolution.
 Dashed lines are square-root fits of the numerical data, while the
 solid lines denotes the analytical calculations.}
 \end{figure}

 As characteristic values for the mean accuracy of single pulses,
 we choose $\delta\phi_0$ in the range of $10^{-6}$ to $10^{-4}$,
 which should be technologically feasible. As one can see in the
 double logarithmic plots of Fig.~\ref{bbfehler}, the numerically
 determined evolutions follow the analytically expected square-root
 type random walk behavior.

 \subsubsection{Random walk due to both inaccurate bb-pulses and
 bfl-noise}

 We now combine our imperfect bang-bang pulse operations with our
 former bfl-noise signal to discuss the applicability of our
 control scheme when ``realistic'' pulse generators are used. As
 before, we calculate the rms deviations at $t_0=\tau_{\rm Sys}$
 by averaging over $10^3$ realizations.  The bfl-parameters are
 those used previously: a coupling strength $\alpha=0.1$ and an
 average switching time $\tau_{\rm bfl}=0.01 \tau_{\rm Sys}$.
 However, with the aim of determining the optimal bang-bang
 protocol in the presence of pulse imperfections, we now consider
 different pulse separation times $\tau_{\rm bb}/\tau_{\rm Sys}$
 ranging from $10^{-5}$ to $10^{-2}$.\\
~\\
 \begin{figure}[h]
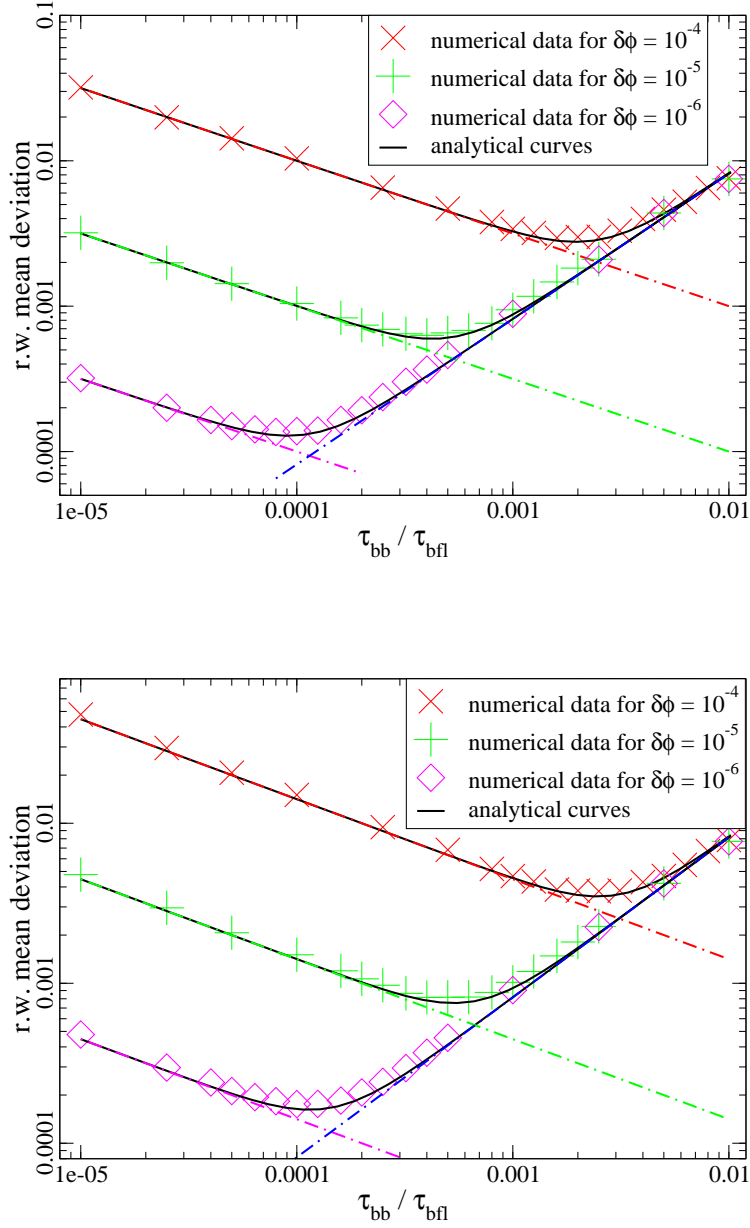

 \includegraphics[width=0.6\columnwidth]{bb1bflplot.eps}\\
~~\\
~~\\
 \includegraphics[width=0.6\columnwidth]{bb2bflplot.eps}
 \caption{\label{bbbflplot}Plot of the Bloch sphere rms deviations
 received by one-/two-dimensional inaccurately pulsed bang-bang
 compensation of the typical bfl-perturbation. Dashed lines describe
 the aberrances for pure faulty bang-bang (i.e.~without bfl-noise),
 respectively the exactly compensated bfl-case (see
 Fig.~\ref{devevolution}), while solid lines denotes the deviations
 calculated by random walk analysis.}
 \end{figure}

 We assume that the errors induced by the bfl and those induced by
 the pulse generator are statistically independent, and thus we sum
 together both sets of induced deviations in the usual rms-fashion.
 In comparison to the case of ideal bang-bang pulses,
 Equ.~(\ref{sigmabb}), we find here the average total deviations
 induced by both  bfl telegraph noise and imperfect bang-bang
 pulses to be:
 \begin{eqnarray}
 \label{dsigmatot1}
 \Delta\sigma_{\rm tot}^{1d} & = & \hspace{4mm}
 \sqrt{\Delta\sigma_{\rm bfl}^2
 \hspace{2mm} +\hspace{2mm} {\Delta\sigma_{\rm bb}^{1d}}^2} \nonumber \\
 & = & \sqrt{ \frac{1}{2} N_{\rm bfl} \alpha^2 \tau_{\rm bb}^2
 + N_{\rm bb} \delta\phi_0^2 } \\
 & = & \sqrt{ \frac{1}{2} \alpha^2 \tau_{\rm bb}^2 \frac{t_0}{\tau_{\rm
 bfl}} + \delta\phi_0^2 \frac{t_0}{\tau_{\rm bb}}} \nonumber
 \end{eqnarray}
 in the one-dimensional case where imperfect pulses only impart
 phase errors (due to imprecise pulse duration), and
 \begin{eqnarray}
 \label{dsigmatot2}
 \Delta\sigma_{\rm tot}^{2d} & = & \hspace{4mm}
 \sqrt{\Delta\sigma_{\rm bfl}^2 \hspace{2mm} +\hspace{2mm}
 {\Delta\sigma_{\rm bb}^{2d}}^2} \nonumber \\
 & = & \sqrt{ \frac{1}{2} N_{\rm bfl} \alpha^2 \tau_{\rm bb}^2
 + 2 N_{\rm bb} \delta\phi_0^2 } \\
 & = & \sqrt{ \frac{1}{2} \alpha^2 \tau_{\rm bb}^2
 \frac{t_0}{\tau_{\rm bfl}} + 2 \delta\phi_0^2 \frac{t_0}{\tau_{\rm bb}}}
 \nonumber
 \end{eqnarray}
 in the two-dimensional case when imperfect pulses impart both
 phase and relaxation/excitation errors (due to imprecision in the
 pulses' polarization axes).

 As Fig.~\ref{bbbflplot} demonstrates, we observe a very good
 agreement between our numerical and analytical results. Such data
 make it possible to determine an optimal bang-bang separation time
 $\tau_{\rm bb}^{\rm opt}$.  Specifically, this optimum can be
 derived by calculating the zero value of the derivative of Equs.
 (\ref{dsigmatot1}) and (\ref{dsigmatot2}) with respect to
 $\tau_{\rm bb}$.  We therefore conclude that the optimal period
 between bang-bang pulses is:
 \begin{eqnarray}
 \label{optaubb1} \tau_{\rm bb}^{1d} & = & \sqrt[3]{
 \tau_{\rm bfl} \frac{\delta\phi_0^2}{\alpha^2}}
 \end{eqnarray}
 for the one-dimensional case and
 \begin{eqnarray}
 \label{optaubb2} \tau_{\rm bb}^{2d} & = & \sqrt[3]{2
 \tau_{\rm bfl} \frac{\delta\phi_0^2}{\alpha^2}}
 \end{eqnarray}
 for the two-dimensional case. These optimal times respectively
 correspond to minimized variances at $t_0=\tau_{\rm Sys}$ of
 \begin{eqnarray}
 \Delta\sigma_{\rm opt}^{1d}  & = & \sqrt{\frac{1}{2}+1} \frac{\alpha^{1/3}
 \delta\phi_0^{2/3}}{\tau_{\rm bfl}^{1/6}} \sqrt{t_0}
 \end{eqnarray}
 for the one-dimensional case of only imprecise pulse durations and
 \begin{eqnarray}
 \Delta\sigma_{\rm opt}^{2d}  & = & \sqrt{2^{-1/3}+2^{2/3}}
 \frac{\alpha^{1/3} \delta\phi_0^{2/3}}{\tau_{\rm bfl}^{1/6}} \sqrt{t_0}
 \end{eqnarray}
 for the two-dimensional case of both imprecise pulse durations and
 polarization axes.

 \section{Conclusion and remarks}

 We have investigated the qubit errors that arise from the noise
 generated by a {\it single} bistable fluctuator (bfl) in its
 semiclassical limit, where it behaves as a telegraph noise source.
 We numerically integrated a corresponding stochastic Schr\"odinger
 equation, Equ.~(\ref{ssesol}), as well as analytically solved (in
 the long-time limit) appropriate random walk models. As a
 characteristic measure of the resulting dephasing and relaxation
 effects, we used the rms deviation of noisy evolutions compared to
 noiseless ones. To suppress the effects of this noise, we
 presented a bang-bang pulse sequence analogous to the familiar
 spin-echo method.   We claimed this pulse sequence to be capable
 of refocusing most of the bfl-noise induced aberrations. Both in
 the case without bang-bang control and the case with it, there was
 excellent agreement between our numerical and analytical results
 on the relevant intermediate time scales (\textit{i.e.}, times
 after a short initial phase where deviations grow linearly instead
 of as a square-root in time, but before the qubit becomes totally
 decohered).

 In particular, we confirmed our preliminary qualitative picture
 that bang-bang control works as a high pass filter, suppressing
 qubit errors by a factor $\mu \tau_{\rm bfl} / \tau_{\rm bb}$ that
 is directly proportional to the ratio of the mean bfl switching
 time and the period between bang-bang pulses. The numerically and
 analytically calculated constants of proportionality $\mu_{\rm
 numerical}\approx 1.679$ and $\mu_{\rm analytical}=\sqrt{5/2}\simeq
 1.581$ also matched to good accuracy. These results imply that the
 bang-bang procedure is an appropriate remedy against the
 $1/f$-noise that often is seen in solid-state environments.  This
 is because bang-bang control exhibits maximal suppression of bfl
 telegraph noise, and $1/f$ noise generally arises from an ensemble
 of bfls. Finally, one has to be aware that also the static 
$\sigma_{\rm z}$-term of the Hamiltonian is averaged out, and this 
generally reduces the degree of control on the qubit. But this is
only a technical constraint, as one could imagine interchanging
two different types of bang-bang pulses (e.g.~along the $x$ and the
$y$-axis respectively) to admit corresponding quantum-gate operations.

 We previously presented this basic idea in a short paper
 \cite{Gutmann03}. The present work extends that short paper by
 treating the effects of different types on non-ideal bang-bang
 pulses. Moreover, the analytical random walk method is outlined in
 much more detail, as this method should also be applicable to
 other problems that are difficult to treat in a master equation
 approach.

 Meanwhile, several other extensions of Ref.~\cite{Gutmann03} have
 been proposed by other research groups.  Ref.~\cite{Altshuler}
 includes a larger number of fluctuators, described as
 semiclassical noise sources, but restricts itself to a single
 spin-echo cycle. Ref.~\cite{Averin} analyzes extensively the
 importance of higher, non-Gaussian cumulants and memory effects
 and arrives at a number of analytical results, but it does not
 treat the option of refocusing. Ref.~\cite{FV03} treats a full
 microscopic model and compares different variations of the
 bang-bang pulse sequence. Ref.~\cite{FAMP03} also treats a full
 microscopic model with potentially many fluctuators using a
 Lindblad-type approach and covers a wide range of ratios between
 the fluctuator and bang-bang pulse time scales.  One of its main
 conclusions is that a Zeno effect is found in a parameter regime
 not covered by our work.  Note that all of these other extensions
 of our work treat only the case of ideal bang-bang pulses.

 \begin{acknowledgements}
 We are indebted to G.~Falci, I.~Goychuk, T.P.~Orlando and J.~von
 Delft for helpful discussions. Our special thanks goes to A.~Kaeck
 for pointing out the capability of our random walk model to
 describe deviations due to imperfect pulsing. HG and FKW also
 thank T.P.~Orlando for his great hospitality during their stay at
 MIT. WMK gratefully acknowledges fellowship support from the
 Fannie and John Hertz Foundation. This work was supported by a
 DAAD-NSF travel grant and by the National Security Agency (NSA) and
Advanced Research and Development Activity (ARDO) under 
Army Research Office (ARO) contract P-43385-PH-QC as well as
 by the DFG through SFB 631.
 \end{acknowledgements}

 \appendix
 \section{} We shall now connect the model of a {\it single} bfl
 as a telegraph noise source to a microscopic Hamiltonian.  We
 start with the conventional Hamiltonian model of a single bfl,
 \textit{e.g.}, \cite{PFFF02,PFF03,GMB02}.  The original qubit is
 influenced by noise from another qubit, the bfl, which itself is
 coupled to a thermal environment by a bilinear spin-boson type
 interaction:
 \begin{eqnarray}
 H & = &  H_{\rm qubit} + H_{\rm qubit,bfl} + H_{\rm bfl} + H_{\rm
 bfl,env} +
 H_{\rm env} .
 \end{eqnarray}
 where
 \begin{eqnarray}
 H_{\rm qubit} & = & \hbar \epsilon_{\rm q} \hat{\sigma}_{\rm z}^{\rm q}
 + \hbar \Delta_{\rm q} \hat{\sigma}_{\rm x}^{\rm q} \\
 H_{\rm qubit,bfl} & = & \hbar \alpha \hat{\sigma}_{\rm z}^{\rm q}
 \hat{\sigma}_{\rm z}^{\rm bfl} \\
 H_{\rm bfl} & = & \hbar \epsilon_{\rm bfl} \hat{\sigma}_{\rm z}^{\rm
 bfl} + \hbar \Delta_{\rm bfl} \hat{\sigma}_{\rm x}^{\rm bfl} \\
 H_{\rm bfl,env} & = & \hbar \lambda\hat{\sigma}_{\rm z}^{\rm bfl} \sum_j
 \left( \hat{a}_j^\dagger + \hat{a}_j \right) \\
 H_{\rm env} & = & \hbar \sum_j \omega_j \left( \hat{a}_j^\dagger
 \hat{a}_j + \id/2 \right)
 \end{eqnarray}
 The scalar $\alpha$ denotes the coupling strength between the
 original qubit and the bfl, while the scalar $\lambda$ indicates
 the influence of the environmental heat bath on the bfl.

 It is not obvious how to treat such a combined open quantum system
 \cite{PFFF02,GMB02}. The common approach of deriving a master
 equation for the reduced qubit system does not work, as it is not
 clear how to introduce an open quantum system ``bfl'' as the
 environment. Gassmann, \textit{et al.} present four alternative
 approaches \cite{GMB02}.  Their first approach is to derive a
 standard Markovian master equation for the combined open system
 ``qubit + bfl'' and trace out the parameters of the bfl
 afterwards.  Their second approach is to consider the qubit as
 influenced by an effective bfl-bath environment by use of an
 Markovian and secular approximation in the limit of small
 $\alpha$.  Their third approach, which is both the most general
 and the most complicated, is to deduce a master equation by
 applying a non-Markovian weak-coupling perturbation ansatz in
 second order in $\alpha$.

 For our investigations, we prefer this last and most general
 approach: a stochastic treatment employing an appropriate randomly
 changing bfl-noise Hamiltonian term (compare also to
 \cite{Altshuler}). This choice is not only because of practical
 reasons (to make our numerics feasible), but also due to empirical
 considerations (see \cite{Harlingen,ZCC97}, where characteristics
 of telegraph noise were observed and attributed to bfls).  Hence,
 we restrict our analysis to the limit $\lambda \gg \alpha$,
 \textit{i.e.}, the limit where the coupling of the bfl to the
 external environment is much larger than interaction between the
 bfl and the qubit. For convenience, we assume the bfl is in its
 high-temperature limit. (Note that this does not necessarily mean
 the qubit is also in a high-temperature regime for the qubit's
 energy scale might be much larger than that of the bfl.)  We thus
 assume the bfl behaves like a classical (\textit{i.e.}, decohered)
 noise source, and we specifically describe the bfl's influence on
 the qubit with the following stochastic Hamiltonian:
 \begin{eqnarray}
 H_{\rm qubit,bfl} \stackrel{\rm semicl.}{\longrightarrow} H_{\rm
   qubit,bfl}^{\rm  noise} (t) & = & \hbar \alpha~\hat{\sigma}_{\rm
 z}^{\rm q}~\xi_{\rm bfl}(t) .
 \end{eqnarray}
 In the equation above, $\xi_{\rm bfl}(t)$ is a random function of
 time representing the switching of the $\sigma_{\rm z}^{\rm
 bfl}$-value between $\pm 1$.  In our high-temperature limit, we
 assume $\xi_{\rm bfl}(t)$ has symmetrical Poissonian statistics
 (\textit{i.e.}, the probabilities of the bfl switching from $+1$
 to $-1$ and from $-1$ to $+1$ are equal and constant over time).
 Such a symmetrical random process is readily described by just one
 parameter: the typical time separation $\tau_{\rm bfl}$ between
 two bfl flips (see Fig.~\ref{bflnoisesignal}).

 The high temperature limit is not a crucial constraint.  Treating
 the strongly thermally coupled bfl in an intermediate temperature
 regime would only result in some asymmetrically switching
 $\xi_{\rm bfl}(t)$. The typical switching times time $\tau_{\rm
 bfl}^{\uparrow,\downarrow}$ for switching the bfl up and down respectively
satisfy the detailed balance relation $\frac{\tau_{\rm
 bfl}^\downarrow}{\tau_{\rm bfl}^\uparrow}=e^{-\delta E_{\rm bfl}/k_b T}$,
where 
 $\delta E_{\rm bfl}$ 
 denotes the energy separation of the two bfl-states, and $T$ the 
 temperature of the heat bath which drives the switching of the bfl. 
 The microscopic structure of the rates depends on details of the experiment. 
 Typically, they will be golden rule rates containing the density of states
 of the heat bath and the matrix element of its coupling to the bfl. 
 If that bath is made of harmonic oscillators with an ohmic spectral density, 
 we e.g.\ expect switching rater $\left(\tau_{\rm bfl}^{\uparrow,\downarrow}\right)^{-1}
 =\pm\frac{\alpha_0 \delta E_{\rm
 bfl}}{e^{\pm\delta E_{\rm bfl}/k_b T}-1} $, where $\alpha_0$ is the 
dimensionless
 coupling strength to the Ohmic bath.
 This would essentially only lead to an
 additional drift of the qubit state, \textit{i.e.}, a random walk
 with a nonzero average value.  Neither our analytical results nor
 our conclusions would otherwise change qualitatively.  In fact,
 assuming the the bang-bang pulse cycles are sufficiently short
 relative both the typical $+1$ to $-1$ and $-1$ to $+1$ switching
 times of the bfl, bang-bang suppression of the bfl noise should
 not be diminished at all by the bfl's asymmetrical switching. We
 therefore obtain Equ.~(\ref{Hstocheq}) as our starting point of
 the bfl-perturbed qubit dynamics.

 \end{document}